\documentclass{article}

\usepackage{fullpage,graphicx,psfrag,amsmath,amssymb,color,pdfpages,dsfont,float,comment,tabularx,subcaption,enumitem,placeins,multirow}
\usepackage[ruled,vlined]{algorithm2e}
\usepackage{siunitx}
\usepackage{booktabs}
\usepackage{authblk}
\usepackage{cite}
\usepackage{url}

\usepackage[hidelinks]{hyperref}
\usepackage{orcidlink}

\title{Likelihood-based anomaly detection \\ in preferential attachment networks}

\author[1]{Qiu Liang\, \orcidlink{0009-0001-7530-7256}}
\author[1]{Remco van der Hofstad\,\orcidlink{0000-0003-1331-9697}}
\author[1]{Nelly Litvak\,\orcidlink{0000-0002-6750-3484}}
\affil[1]{Department of Mathematics and Computer Science, Eindhoven University of Technology, Eindhoven, The Netherlands}

\date{}
\begin{document}

\maketitle

\begin{abstract}
Preferential attachment (PA) network is a widely used model for capturing the growth dynamics of real-world networks, in which newly arriving vertices are more likely to connect to existing vertices with higher degrees. In this paper, we consider a setting in which an anomalous vertex appears at some time point and receives edges with an additional attachment advantage governed by a parameter $\beta$, while ordinary vertices continue to follow the PA mechanism with parameter $\delta$. Detecting such anomalies is challenging due to the high variability in degree growth in PA networks and the limited information available when the anomaly arrives late in the network evolution. We propose an iterative parameter estimation procedure together with a likelihood-based detection framework. Simulation results show that the proposed procedure provides accurate estimation of the parameters 
$\beta$ and $\delta$. Detection performance depends on the time of anomaly occurrence: anomalies arising at midway stages of the network evolution are detected most reliably, whereas very early and late anomalies remain challenging, particularly when the parameter $\beta$ is small.
\end{abstract}


\section{Introduction}

Detecting anomalies in dynamic networks is a vital and challenging problem in modern network science \cite{Ranshous2015Survey}, with critical applications including financial fraud detection \cite{Akoglu2015detection}, social media \cite{Savage2014SocialNetwork} and information diffusion systems \cite{Soroush2018SpreadNews}. Existing approaches can be broadly categorized into machine learning methods, which learn temporal graph representations~\cite{LiuDectectionTransformer, YuNetwalk2018} or anomaly scores~\cite{ZhengAddGraph2019} directly from data, and statistical methods, which model network evolution and flag significant structural deviations~\cite{heard2010bayesian, Peel2015Baysian}. Despite their successful empirical results, machine learning methods often require large amounts of training data~\cite{LiuDectectionTransformer, YuNetwalk2018,ZhengAddGraph2019}, and statistical methods impose strong assumptions about the nature of structural changes in the network~\cite{heard2010bayesian, Peel2015Baysian}.

In contrast, in this work, we focus on detecting anomalies in networks generated by the Preferential Attachment model, a well-established generative model for real-world networks such as social~\cite{barabasi2002evolution} and citation networks~\cite{barabasi1999emergence}. PA networks are characterized by distinctive properties such as the ``rich-get-richer'' phenomenon and power-law degree distributions, making deviations from the PA mechanism crucial for identifying malicious behavior and structural shifts. Existing research on anomalies in PA networks has primarily focused on global structural changes, such as change-point detection associated with variations in the additive fitness parameter~\cite{banerjee2023fluctuation, kay2023detecting, bhamidi2018longrange, cirkovic2022likelihood, du2025Likelihood, Kaddouri2026Likelihood}. Related work has also examined anomalies from a distributional perspective, for example through stability analyses of the Barab\'{a}si--Albert model that characterize the asymptotic behavior and robustness of the limiting degree distribution under small perturbations~\cite{Ruiz2019LyapunovAnomalyDetection}. 

We instead consider a different setting in which the fitness parameter $\delta$ remains fixed, while a specific vertex exhibits abnormal behavior by attracting an unusually large number of new edges through an additional parameter $\beta$. This type of anomaly alters the local generative mechanism of the PA network without changing the global model parameters. Detecting such anomalies is challenging for several reasons. The inherent high variability in degree growth under PA networks makes it hard to distinguish an anomalous vertex from the usual high-degree vertices present in PA models. Additionally, accurately localizing the anomalous vertex is particularly difficult when the anomaly occurs late in the network evolution, or when the anomalous signal is weak. Moreover, because the parameters $\beta$ and $\delta$ are not known a priori and must be estimated from the observed data under the assumed model, the detection procedure requires fitting these parameters for each candidate vertex. This can become computationally demanding as the network size grows, motivating the need to limit the candidate search range.

To address these challenges, we propose a likelihood-based detection procedure for PA networks. Our approach first estimates the network parameters using maximum likelihood estimation (MLE) based on simulated networks, and then formulates a hypothesis-testing framework to determine whether an anomaly is present or not. To reduce computational complexity, we further restrict the search space by identifying a small set of candidate vertices based on degree-related characteristics and abnormal degree growth patterns.

Building on the anomaly model introduced in our earlier work \cite{Qiu2025Degrees}, this paper makes the following contributions:
\begin{enumerate}[label=(\arabic*)]
    \item We propose an iterative procedure for estimating the model parameters $\beta$ and $\delta$ in the presence of an anomalous vertex.
    \item We propose a likelihood-based framework for detecting vertex-level anomalies in PA networks, which accommodates both known and unknown $\delta$.
    \item We conduct a systematic simulation study across early, midway, and late anomaly regimes, demonstrating that anomalies occurring at intermediate stages of the network evolution are detected most reliably, whereas very early and late anomalies with small anomaly strength $\beta$ remain challenging to detect.
\end{enumerate}

The rest of this paper is organized as follows. In Section 2, we provide the construction of the PA network incorporating an anomaly, along with different types of the anomaly. We next derive the MLEs of network parameters in Section 3. In Section 4, we propose the hypothesis-testing framework for detecting the anomaly. The performances of estimation and anomaly detection are evaluated through numerical experiments in Section 5. Section 6 discusses the limitations of our approach, the broader implications, and the comparison with previous studies, while Section 7 concludes the paper and outlines potential extensions.

\section{PA network with an anomaly}
\label{sec:model}

We follow the formulation of the preferential attachment model given in \cite[Chapter~8]{Hofstad_2016}. This construction is equivalent to the models studied in \cite{BerBorChaSab14, kay2023detecting, BerBorChaSab05}. The network starts with a single vertex and $m$ self-loops, where $m \geq 1$. 
At each time step $t > 1$, a new vertex $v_t$ arrives with $m$ edges. 
These edges are added sequentially: for $j \in [1,m]$, the graph $G_{t,j}$ is obtained from $G_{t,j-1}$ by attaching one edge of $v_t$ to an existing vertex. We denote the final graph at time $t$ by $G_t = G_{t,m}$. Conditional on $G_{t,j-1}$, the probability that the $j$-th edge of $v_t$ attaches to an existing vertex $v_i \in \{v_1, \dots, v_{t-1}\}$ is given by

\begin{equation}
    \label{rule-1}
    P(v_{t,j} \to v_i \mid  G_{t,j-1}) = \frac{D_i(t,j-1) + \delta}{2m(t-1)+(t-1)\delta + (j-1)}, \quad t>1,
\end{equation}
where $D_i(t,j-1)$ represents the degree of vertex $v_i$ in $G_{t,j-1}$, and $\delta > -m$.

In our previous work~\cite{Qiu2025Degrees}, we extended this model by introducing an \emph{anomaly} that occurs at some time $\tau$ with $ 1< \tau < t$, after which the attachment rule is modified. We denote the anomaly as $v_\tau$. The attachment rule  {\em at step $t$} is then as follows:

\begin{enumerate}
    \item[(I)] For $ t<\tau $, $G_t$ evolves according to \eqref{rule-1}.
    \item[(II)] For $ 1 < \tau \leq t$, the dynamics are updated as 
\begin{equation}
\label{rule-2}
P(v_{t,j} \to v_i \mid G_{t,j-1})=
\begin{cases}
\displaystyle
 \frac{D_{i}(t,j-1)+\delta}{(t-1)(2 m+ \beta + \delta)+ j-1} & \text{ if } i \neq \tau ,\\
 \displaystyle\frac{(t-1) \beta+ D_{\tau}(t,j-1)+\delta}{(t-1)(2 m+ \beta + \delta)+ j-1
 } & \text{ if } i = \tau ,
\end{cases}
\end{equation}
\end{enumerate}
where $\beta>0$. To streamline the notation, we summarize the two growth mechanisms introduced above. 
We write $\mathrm{PA}(\delta)$ for the standard PA model 
following rule~\eqref{rule-1}. We denote the PA model incorporating an anomaly by $\mathrm{PA}(\beta,\delta;\tau)$, as described by rule~\eqref{rule-2}.

As indicated by the rule~\eqref{rule-2}, the arrival time $\tau$ of the anomalous vertex plays a vital role in determining the evolution of the preferential attachment network with an anomaly. Depending on how $\tau$ scales with the network size $t$, we distinguish three regimes:
\begin{enumerate}[label = (\arabic*)]
\item \emph{Early anomaly}: $\tau = t^\gamma$, $\gamma \in [0,1)$;
\item \emph{Midway anomaly}: $\tau = \gamma t$, $\gamma \in (0,1)$;
\item \emph{Late anomaly}: $\tau = t - t^\gamma$, $\gamma \in (0,1)$.
\end{enumerate}
In our previous work \cite{Qiu2025Degrees}, we found that these regimes lead to markedly different asymptotic behaviors of the degree distribution. When the anomaly occurs late, its influence on the global network structure is negligible, and the degree distribution coincides with that of a standard PA network, exhibiting a power-law exponent $3 + \delta/m$. In the midway anomaly case, the impact of the anomaly remains localized: vertices arriving after time $\tau$ predominantly have bounded degrees, while those born before $\tau$ follow a power-law distribution with exponent $3 + \delta/m$, multiplied by an additional factor $\gamma^{\frac{\beta}{2m+\beta+\delta}}$. By contrast, an early anomaly substantially alters the network evolution, leading to a modified power-law exponent $3 + \delta/m + \beta/m$ and the anomaly having degree of order $t$.

The primary goal of this study is to identify the presence of an anomaly within the PA network. The task naturally decomposes into two subproblems: parameter estimation and anomaly detection.

\section{Maximum likelihood estimation for parameters}
\label{sec: estimation}

In this section, we develop a procedure to estimate the parameters of the $\mathrm{PA}(\beta,\delta;\tau)$ model. In our empirical framework, we treat model parameters as unknown and estimate them from the observed network. These estimates forms the basis for the anomaly detection introduced in the next section.

\subsection{Parameter estimation in PA models}

We begin by introducing the estimation framework for the standard model 
$\mathrm{PA}(\delta)$. The estimation of $ \delta$ has been well studied in the literature, for example, \cite{Gao2017MLE} proposes an MLE approach and establishes its consistency.

Given the graph sequence $ ( G_s )_{s = 2}^{t} $, let $u_{s,j}$ denote the vertex that receives the $j$-th incoming edge at time $s$, and let $D_{u_{s,j}}(s,j)$ be the degree of vertex $u_{s,j}$ in $G_{s,j}$, where $ j\in [1,m]$. Then the likelihood function of observing the graph sequence $ ( G_s )_{s = 2}^{t} $ is
\begin{equation}
    \label{eq: L_0, likelihood function without anomaly}
     L(\delta \mid (G_s)_{s = 2}^{t} ) = \prod_{s=2}^{t} \prod_{j=1}^{m}\frac{D_{u_{s,j}}(s,j-1) + \delta}{(s - 1)(2m+\delta) + j-1}.
\end{equation}
The corresponding log-likelihood function is then given by 
\begin{equation}
    \label{eq:log L_0}
    \log L(\delta \mid (G_s)_{s = 2}^{t} ) = \sum_{s=2}^{t} \sum_{j=1}^{m}\left\{ \log \left[ D_{u_{s,j}}(s,j-1) + \delta \right] - \log\left[ (s - 1)(2m+\delta) + j-1 \right]\right\}.
\end{equation}
Taking the derivative with respect to $\delta$, we obtain
\begin{equation}
    \frac{\partial \log L(\delta \mid (G_s)_{s = 2}^{t})}{\partial \delta} = \sum_{s = 2}^{t}\sum_{j=1}^{m} \left[ \frac{1}{D_{u_{s,j}}(s,j-1) + \delta} - \frac{s -1}{(s - 1)(2m+\delta) + j-1}\right].
\end{equation}
The MLE of $\delta$ is the solution of 
\begin{equation}
 \frac{\partial \log L(\delta \mid (G_s)_{s = 2}^{t})}{\partial \delta} = 0.
\end{equation}

This estimator of $\delta$ is consistent and asymptotically normal distributed, providing a solid basis for statistical inference, as established in~\cite{Gao2017MLE}. In the next subsection, we will generalize this method to our model with an anomaly.

\subsection{Parameters estimation in PA models with an anomaly}

Inspired by \cite{Gao2017MLE}, we adopt a similar estimation framework to estimate the parameters $\beta$, $\delta$ and $\tau$ in our model.

\subsubsection{Construction of the likelihood function}

Assume that the anomaly occurs at time $ \tau>1$. Then the likelihood function becomes 
\begin{equation}
\label{eq: L_1, likelihood function with anomaly }
    L(\beta, \delta, \tau \mid (G_s)_{s = 2}^{t} ) = \prod_{s_1=2}^{\tau} \prod_{j=1}^{m}\frac{D_{u_{s,j}}(s_1,j-1) + \delta}{(s_1 - 1)(2m+\delta) + j-1}\prod_{s_2=\tau+1}^{t}\prod_{j=1}^{m}\frac{S(s_2,j-1)}{(s_2-1)(2m+\beta + \delta)+j-1},
\end{equation}
where 
$$ S(s_2,j-1) = \mathds{1}_{\left\{ u_{s,j} \neq  v_\tau \right\}}(D_{u_{s,j}}(s_2,j-1) + \delta) + \mathds{1}_{\left\{ u_{s,j} = v_\tau \right\}}\left[ D_\tau(s_2,j-1) + \delta + (s_2-1)\beta \right].$$
As a result, the log-likelihood function is given by
\begin{align}
\label{eq: log L_1:Log-likelihood function with anomaly}
\nonumber
        \log L(\delta, \beta, \tau \mid (G_s)_{s = 2}^{t} ) 
        &= \sum_{s_1=2}^{\tau} \sum_{j=1}^{m} 
        \log \left[ D_{u_{s,j}}(s_1,j-1) + \delta \right]
        - \sum_{s_1=2}^{\tau} \sum_{j=1}^{m} 
        \log \left[ (s_1 - 1)(2m+\delta) + j-1 \right] \\
        &\quad + \sum_{s_2=\tau +1}^{t} \sum_{j=1}^{m} 
        \log S(s_2,j-1)
        - \sum_{s_2=\tau +1}^{t} \sum_{j=1}^{m} 
        \log \left[ (s_2 - 1)(2m + \beta + \delta) + j-1 \right].
\end{align}

In the standard PA network, all vertices follow the same attachment mechanism, and the resulting likelihood function \eqref{eq:log L_0} is typically smooth and well behaved, as shown in \cite{Gao2017MLE}. In contrast, our model introduces an anomalous vertex whose attachment dynamics differ from those of ordinary vertices, resulting in a more complex likelihood function with strong interactions between parameters, particularly between the anomaly strength $\beta$ and the fitness parameter $\delta$. Moreover, since $\tau$ is a discrete parameter, the likelihood is not differentiable with respect to $\tau$.

\begin{figure}
    \centering
    \includegraphics[width=0.75\linewidth]{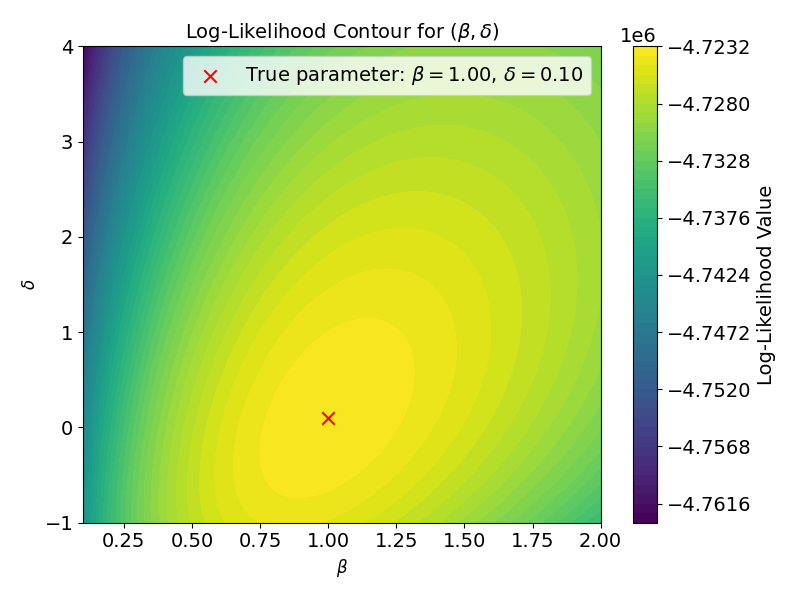}
    \caption{The contour plot of the log-likelihood function evaluated at $ t = 100000, \tau = 20000, m = 5, \beta = 1.0, \delta = 0.1.$}
    \label{fig: contour surface of log L}
\end{figure}

As a result, the Hessian of the likelihood function in \eqref{eq: log L_1:Log-likelihood function with anomaly} does not appear to admit a tractable closed-form expression, making it challenging to characterize the global curvature of the likelihood over the parameter space. To illustrate the behavior of the log-likelihood function, we therefore evaluate it numerically over a grid of plausible parameter values and visualize the resulting contour surface. As shown in Figure~\ref{fig: contour surface of log L}, the log-likelihood exhibits relatively flat regions in the neighborhood of the true parameter values. As a consequence, the maximization problem becomes numerically challenging, since small variations of the parameters induce only minor changes in the likelihood value. Motivated by this observation, we next introduce an estimation strategy to achieve stable convergence despite the flatness of the likelihood surface.

\subsubsection{Estimation procedure}

Joint estimation is necessary to account for the combined effect of the model parameters on the network evolution and to obtain coherent parameter estimates when the anomaly occurrence and strength are unknown.

Let $\boldsymbol{\theta} = (\beta, \delta, \tau)$, and define the log-likelihood function as
\begin{equation}
    \ell(\boldsymbol{\theta}) = \log L(\boldsymbol{\theta} \mid (G_s)_{s = 2}^{t}).
\end{equation}
Then the MLE of the parameter vector 
$\theta$ is given by
\begin{equation}
    \hat{\boldsymbol{\theta}} = \arg\max_{\boldsymbol{\theta} \in \Theta} \ell(\boldsymbol{\theta}),
\end{equation}
where the parameter space $\Theta $ is
\begin{equation}
    \Theta = \mathbb{R}^{+} \times (-m,\infty) \times \{1,2,\dots,t\}.
\end{equation}
In the numerical implementation, we minimize the negative log-likelihood rather than maximizing $\ell$ directly. Since the anomaly location $\tau$ is discrete, we evaluate the likelihood over all possible values of $\tau$ and select the value that minimizes the negative log-likelihood. For the continuous parameters $\delta$ and $\beta$, although their theoretical domains are $(-m, +\infty)$ and $(0, +\infty)$, respectively, we restrict the search space in practice to $\delta \in (-m, a)$ and $\beta \in (0, b)$ for finite constants $a>0$ and $b > 0$, in order to ensure numerical stability and computational efficiency. The joint estimator $\hat{\boldsymbol{\theta}} = (\hat{\beta}, \hat{\delta}, \hat{\tau})$ is obtained by minimizing the negative log-likelihood function over the full parameter space $\Theta$.

Joint optimization in our setting is challenging for several reasons. Specifically, the log-likelihood function is non-differentiable with respect to the anomaly location $\tau$, which is a discrete parameter. Even when conditioning on $\tau$, the score equations for the continuous parameters $\beta$ and $\delta$ do not admit closed-form expressions, indicating that derivative-based optimization methods are impractical. Moreover, the parameter $\beta$ influences only the attachment dynamics of the anomalous vertex, whereas $\delta$ affects the attachment probabilities of all vertices throughout the network evolution. Results obtained via naive joint maximization exhibit a compensation effect between $\beta$ and $\delta$: increases in $\delta$ can partially offset the contribution of the anomaly, leading the optimization to favor solutions with small or near-zero values of $\beta$. As a result, such procedures may produce unstable estimates and underestimate the anomaly strength.

Motivated by these considerations and by the empirical behavior of the likelihood surface discussed earlier, we adopt an iterative joint estimation procedure combined with a multi-start strategy. A key design choice is the use of multiple initial values for $\beta$, rather than for the parameter $\delta$. Employing multiple initial values for $\delta$ does not address this issue, as variations in $\delta$ can further absorb the effect of the anomaly. In contrast, initializing the optimization from a range of values of $\beta$ encourages exploration of different anomaly strengths and reduces the tendency of the estimator to converge to solutions in which $\beta$ is underestimated. For this reason, we adopt a multi-start strategy over $\beta$, while in the subsequent steps we  update $\delta$ and $\beta$ within each iteration.  Joint estimation is carried out by iterating the bounded one-dimensional updates until the change of estimates falls below a prescribed tolerance or the maximum number of iteration is reached. Algorithm~\ref{alg:joint-estimation} summarizes the resulting estimation procedure.

\begin{algorithm}[t]
\caption{Joint Estimation Procedure}
\label{alg:joint-estimation}
\KwData{Initial value sets $\{\beta_{0,1}, \beta_{0,2}, \cdots, \beta_{0,l} \}$ ; tolerance $\text{tol}$; maximum iterations $\text{max\_iter}$; candidate vertices.}
\KwResult{Parameter estimates $\hat{\beta}$, $\hat{\delta}$, and anomaly location $\hat{\tau}$.}

\ForEach{$\beta_{0,i}$ \textnormal{in initial value sets}}{
    \ForEach{vertex $\tau$ \textnormal{in candidate vertices}}{
        
        % Initialize parameters and iteration counter
        $\beta^{(0)} \gets \beta_{0,i}$; \\
        $k \gets 0$; \\[0.5em]

        % Iterative updates
        \While{$k < \text{max\_iter}$ \textnormal{and not converged}}{
            $k \gets k + 1$; \\

            \tcp{Update $\delta$ given current $\beta$}
            $\delta^{(k)} \gets \displaystyle \arg\min_{\delta} -\log L(\beta^{(k-1)}, \delta, \tau)$; \\[0.3em]

            \tcp{Update $\beta$ given current $\delta$}
            $\beta^{(k)} \gets \displaystyle \arg\min_{\beta} -\log L(\beta, \delta^{(k)}, \tau)$; \\[0.3em]

            \tcp{Check convergence}
            Compute $|\beta^{(k)} - \beta^{(k-1)}|$ and $|\delta^{(k)} - \delta^{(k-1)}|$; \\
            \If{$|\beta^{(k)} - \beta^{(k-1)}| < \text{tol}$ \textnormal{and} $|\delta^{(k)} - \delta^{(k-1)}| < \text{tol}$}{
                break; % Convergence reached
            }
        }

        % Store results for this tau
        Compute $\ell_\tau = -\log L(\beta^{(k)}, \delta^{(k)}, \tau)$; \\
        Store $(\beta^{(k)}, \delta^{(k)}, \tau, \ell_\tau)$;
    }
}

% Select final estimate
$(\hat{\beta}, \hat{\delta}, \hat{\tau}) \gets \displaystyle \arg\min \ell_\tau$ \textnormal{over all stored results}. \\
\end{algorithm}

\section{Likelihood-based anomaly detection}
\label{sec: detection}

In this section, we develop a likelihood-based anomaly detection procedure for PA networks. Building on the parameter estimation framework introduced in Section~\ref{sec: estimation}, we assess whether the observed network evolution deviates significantly from the fitted baseline PA model. The goal is to determine whether the data provide statistical evidence for the presence of an anomaly.

\subsection{Detection framework}

We formalize the detection problem as a hypothesis-testing framework. Specifically, we construct a test statistic based on the likelihood function and determine its significance under the null hypothesis. The proposed detection method is inspired by the likelihood-based approach developed in \cite{cirkovic2022likelihood} for change-point detection in PA networks, where the change-point corresponds to a shift in the parameter $\delta$. In contrast, we assume $\delta$ remains fixed and instead detect anomalies arising from an additional attachment mechanism.

We consider the following binary hypotheses:
\begin{align}
H_0 &: \beta = 0,\quad G_t \sim \mathrm{PA}(\delta), \nonumber \\
H_a &: \beta > 0,\quad G_t \sim \mathrm{PA}(\beta,\delta;\tau). \nonumber
\end{align}
We construct the test statistic under the assumption that the parameter $\delta$ is unknown. 
Let $ \hat{\delta}_0$ and $(\hat{\beta}, \hat{\delta}_1, \hat{\tau}) $ denote the parameter estimates obtained under $H_0$ and $H_a$, respectively. The likelihood ratio is defined as the ratio of the maximized likelihoods under
the null and alternative models. Evaluating the likelihoods at the corresponding MLEs yields
\begin{align}
\Lambda
=
\frac{L(\hat{\delta}_0 \mid (G_s)_{s = 2}^{t})}
     {L(\hat{\beta}, \hat{\delta}_1, \hat{\tau} \mid (G_s)_{s = 2}^{t})}.
\end{align}
We consider the statistic $T(G) = -\log \Lambda$, where large values indicate the presence of an anomaly. 
Under the alternative hypothesis, the estimated anomaly location $\hat{\tau}$ is defined as the value of $\tau$ that maximizes $-\log \Lambda$.

Although \cite{cirkovic2022likelihood} provides the asymptotic distribution of the likelihood-based test statistics under the null hypothesis, their results rely on the existence of a consistent estimator for $\delta$. In the present setting, a closed-form estimator for $\delta$ is available under the null model~\cite{Gao2017MLE}, but not when the network contains an anomaly. In the latter case, the estimator lacks a closed-form expression, and its consistency has not yet been theoretically established. Consequently, deriving the limiting distribution of the proposed likelihood ratio test statistic under $H_0$ is analytically intractable, and we instead approximate the null distribution using a simulation-based approach. Specifically, we first estimate the parameters $(\hat{\beta}, \hat{\delta}_1, \hat{\tau})$ using Algorithm~\ref{alg:joint-estimation}, and generate $N$ null graphs with $\hat{\delta}_1$,  then compute $T(G)$ for those null graphs and use the $(1-\alpha)$-quantile of obtained values, as a boundary, $c_{1-\alpha}$, of the critical region. Note that we generate the null graph using $\hat \delta_1$ estimated under $H_a$ rather than $\hat \delta_0$ estimated under $H_0$. This is because when the observed graph contains an anomaly, fitting the null model may cause $\hat{\delta}_0$ to absorb part of the anomalous structure. By estimating $\delta$ under $H_a$, where the anomaly is explicitly represented, $\hat{\delta}_1$ provides a calibration of the attachment mechanism that is less likely to mask the effect of the anomaly. The detection procedure is summarized in Algorithm~\ref{alg:detection_method}.

Because the detection procedure incorporates parameters estimation including $\tau$, the resulting test provides information not only about the presence of the anomaly, but also its location when an anomaly is detected in the network. We evaluate the performance measure for the detection procedure  in Algorithm~\ref{alg:detection_method} by the power of  the test
\begin{align}
1-\mathbb{P}_{H_1}\left( T(G) \leq c_{1-\alpha} \right),
\end{align}
which is estimated in our numerical experiments as the proportion of networks generated under $H_a$ for which the test statistic is larger than $c_{1-\alpha}$.

\begin{algorithm}[h]
\caption{Detection Procedure}
\label{alg:detection_method}
\KwData{ Simulated graph $G$; number of simulation replicates $N$; significance level $\alpha$.}
\KwResult{Anomaly location $\hat{\tau}$, critical value $c_{1-\alpha}$ and rejection decision for $H_0$.}
\BlankLine

\textbf{Step 1: Compute the Test Statistic $T\!\left(G\right)$}\\
\Indp
    \textbf{(a)} Under $H_a$: estimate $(\hat{\beta}_1, \hat{\delta}_1, \hat{\tau}_1)$ using Algorithm~\ref{alg:joint-estimation} and compute the log-likelihood
        $\log L_1\!\left(\hat{\beta}_1,\, \hat{\delta}_1,\, \hat{\tau}_1 \mid G\right)$\;
    \textbf{(b)} Under $H_0$: estimate $\hat{\delta}_0$ and compute the log-likelihood $\log L_0\!\left(\hat{\delta}_0 \mid G\right)$\;
    \textbf{(c)} Compute the likelihood ratio test statistic
    \[
        T\!\left(G\right) = \log L_1\!\left(\hat{\beta}_1,\, \hat{\delta}_1,\, \hat{\tau}_1 \mid G\right) - \log L_0\!\left(\hat{\delta}_0 \mid G\right).
    \]
\Indm
\BlankLine

\textbf{Step 2: Estimate the Critical Value Under $H_0$}\\
\Indp
\For{$i = 1, \dots, N$}{
    Generate a replicate graph $\widetilde{G}^{(i,\hat{\delta}_1)}$ under attachment rule~\eqref{rule-1} with estimated parameter $\hat{\delta}_1$\;
    Repeat steps \textbf{(a)}, \textbf{(b)} and \textbf{(c)} above to compute 
    $T \!\left(\widetilde{G}^{(i,\hat{\delta}_1)}\right)$\;}
    
    Set $c_{1-\alpha}$ to be the empirical $(1-\alpha)$-quantile of $\left\{T \!\left(\widetilde{G}^{(i,\hat{\delta}_1)}\right)\right\}_{i=1}^{N}$\;

\Indm
\BlankLine

\textbf{Step 3: Rejection Decision (Anomaly Detection) and Anomaly Location}\\
\Indp
    Reject $H_0$ in favor of $H_a$ if $T\!\left(G\right) > c_{1-\alpha}$, and set the anomaly location as \[\hat{\tau} =\arg\max_{\hat \tau_1}T(G).\]
\Indm
\end{algorithm}

\subsection{Empirical type-I error}

We now discuss the calibration of the proposed detection procedure and its control of the type-I error. 
The type-I error is defined as the probability of rejecting the null hypothesis that the observed network evolution follows the normal PA mechanism, that is,
\begin{align}
\mathbb{P}_{H_0}\left(T(G) > c_{1-\alpha} \right).
\end{align}
When the distribution of $T(G)$ is known, the probability of type-I error equals exactly to $\alpha$. However, the distribution of $T(G)$ is not available in closed form for our model; instead, $c_{1-\alpha} $ is obtained empirically in {\bf Step 2} of Algorithm~\ref{alg:detection_method}.
The empirical type-I error is then estimated as
\begin{align}
\hat{\alpha} = \frac{1}{N} \sum_{i = 1}^{N} \mathds{1}_{\left\{ T\!\left(G^{(i,\hat{\delta}_1)}\right) > c_{1-\alpha}^{(i)} \right\}},
\end{align}
where $N$ is the number of replicate networks. We investigate whether the empirical type-I error is close to $\alpha$ by performing the steps described in Algorithm~\ref{alg:type-I-error}.

\begin{algorithm}[h]
\caption{Simulation-based Estimation of the Type-I Error}
\label{alg:type-I-error}
\KwData{Parameter $\delta$; network size $t$; significance level $\alpha$; number of simulation replicates $N_1$.}
\KwResult{Estimated type-I error rate at level $\alpha$.}
\BlankLine
\tcp{Step 1: Generate baseline networks}
Generate $N_1$ independent PA networks $\widetilde{G}_{t}^{(1,\delta)}, \dots, \widetilde{G}_{t}^{(N_1,\delta)}$ under the null model with parameter $\delta$\;
\BlankLine
\tcp{Step 2: Test each network}
\Indp
\For{$i = 1, \dots, N_1$}{
    Apply Algorithm~\ref{alg:detection_method} to $\widetilde{G}_{t}^{(i,\delta)}$ to obtain test statistic $T\!\left(\widetilde{G}_{t}^{(i,\delta)}\right)$ and critical value $c_{1-\alpha}^{(i)}$\;

    \lIf{$T\!\left(\widetilde{G}_{t}^{(i,\delta)}\right) > c_{1-\alpha}^{(i)}$}{record a rejection of $H_0$}
}
\Indm
\BlankLine
\tcp{Step 3: Estimate type-I error}
\KwRet{$\hat{\alpha} = \dfrac{1}{N_1} \displaystyle\sum_{i=1}^{N_1} \mathds{1}_{\left\{T\!\left(\widetilde{G}_{t}^{(i,\delta)}\right) > c_{1-\alpha}^{(i)}\right\}}$}
\end{algorithm}

\subsection{Regimes of detectability}

In this subsection, we study how anomaly detectability depends on the model parameters. In particular, we focus on parameter regimes in which the anomalous vertex receives only a limited number of additional edges, making the anomaly difficult to detect. 

\begin{figure}[!htbp]
    \centering

    % early
    \begin{minipage}{0.48\textwidth}
        \centering
        \includegraphics[width=\linewidth]{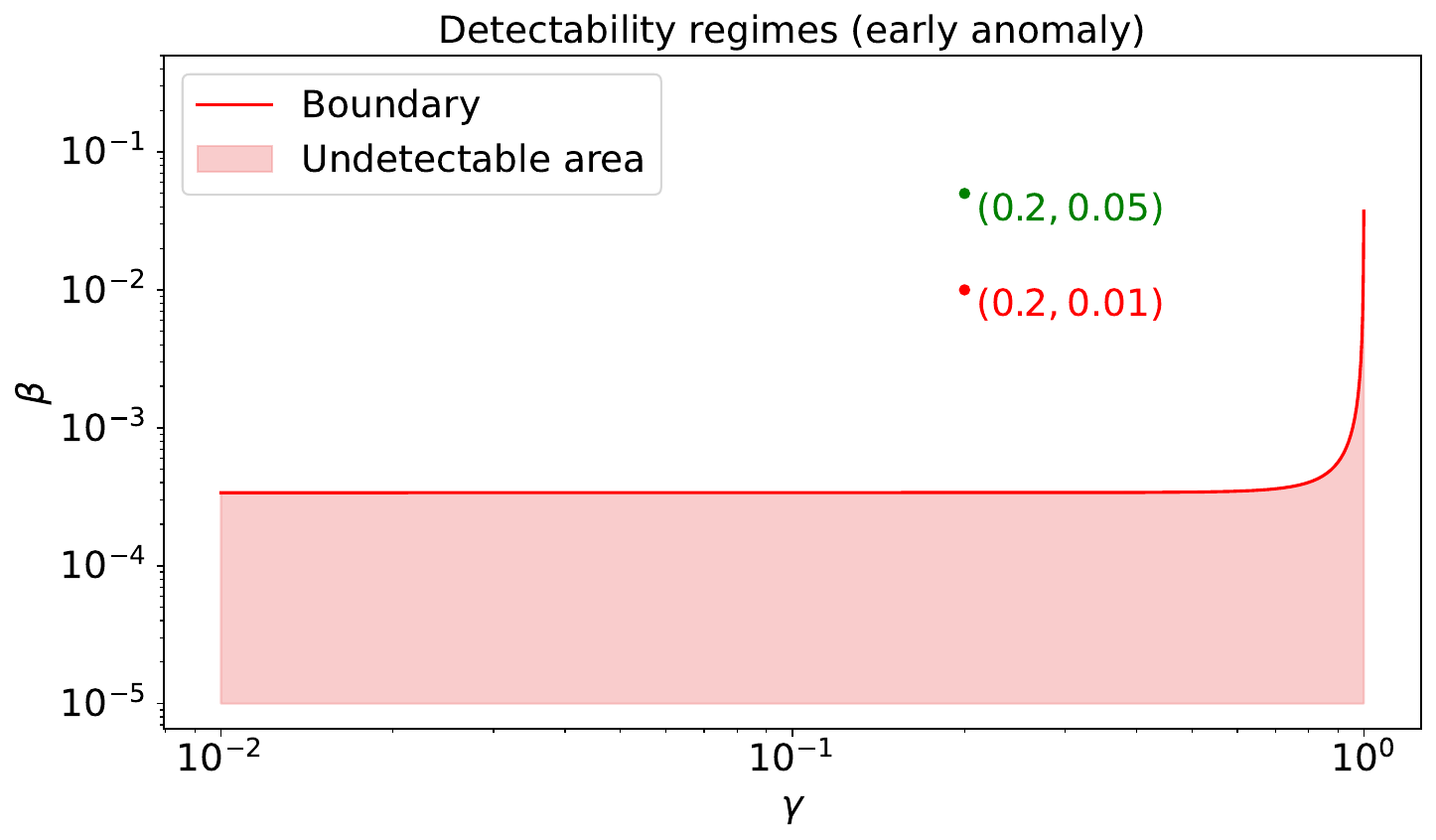}
        \subcaption{$\tau = t^{\gamma}$, $t = 10^{4}$}
    \end{minipage}
    \hfill
    \begin{minipage}{0.48\textwidth}
        \centering
        \includegraphics[width=\linewidth]{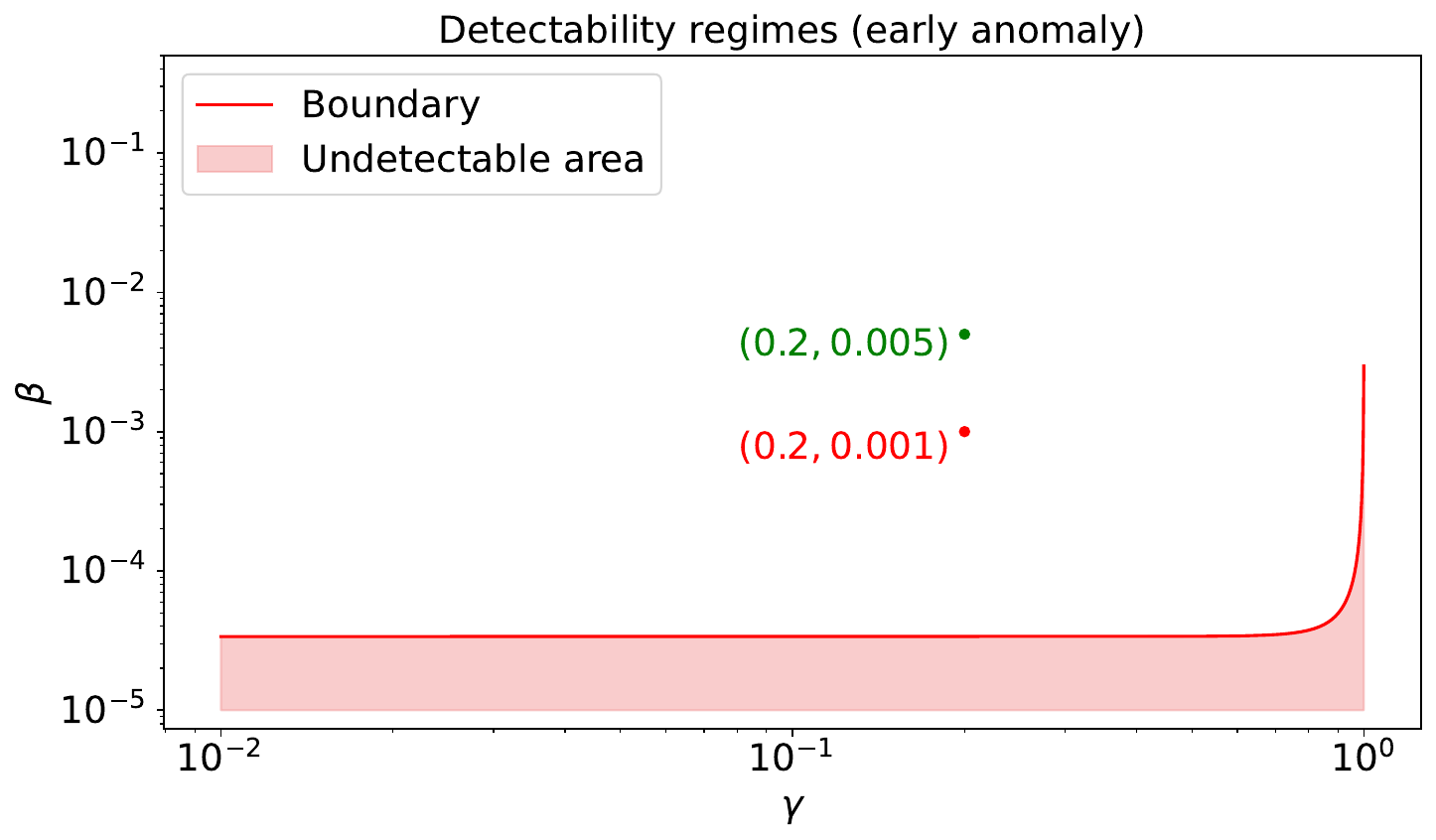}
        \subcaption{$\tau = t^{\gamma}$, $t = 10^{5}$}
    \end{minipage}

    \vspace{0.4cm}

    % midway
    \begin{minipage}{0.48\textwidth}
        \centering
        \includegraphics[width=\linewidth]{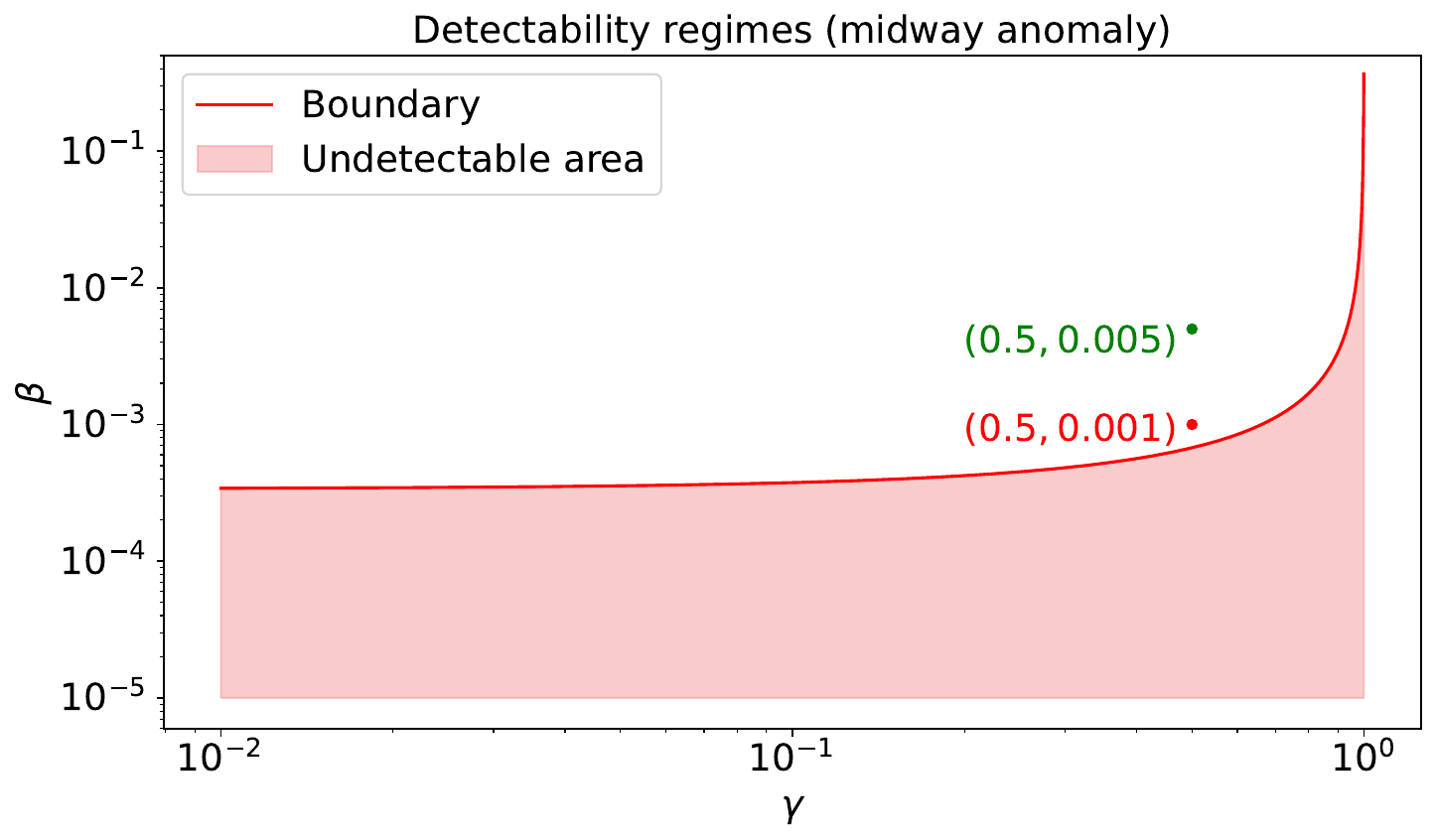}
        \subcaption{$\tau = \gamma t$, $t = 10^{4}$}
    \end{minipage}
    \hfill
    \begin{minipage}{0.48\textwidth}
        \centering
        \includegraphics[width=\linewidth]{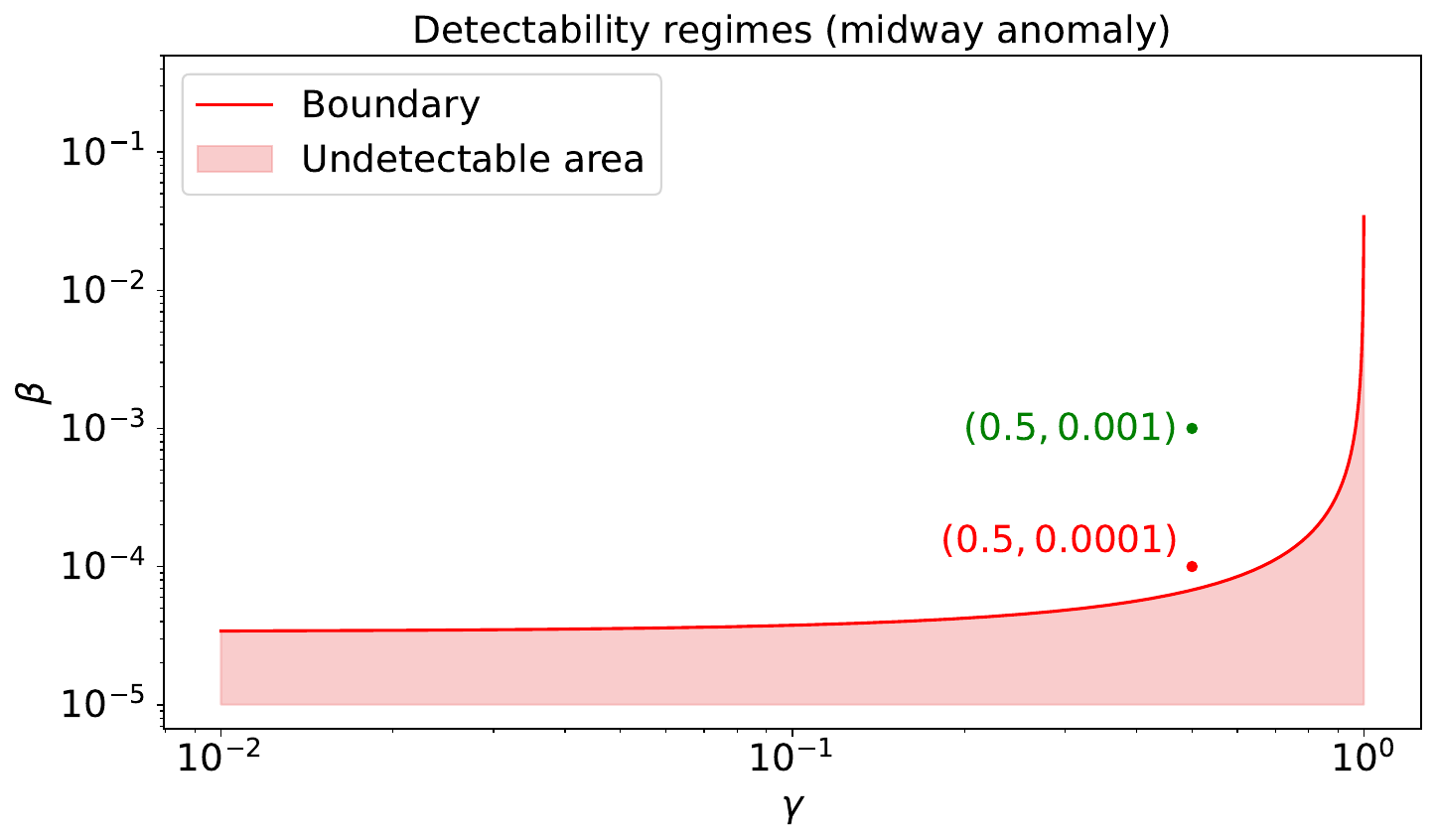}
        \subcaption{$\tau = \gamma t$, $t = 10^{5}$}
    \end{minipage}

    \vspace{0.4cm}

    % late
    \begin{minipage}{0.48\textwidth}
        \centering
        \includegraphics[width=\linewidth]{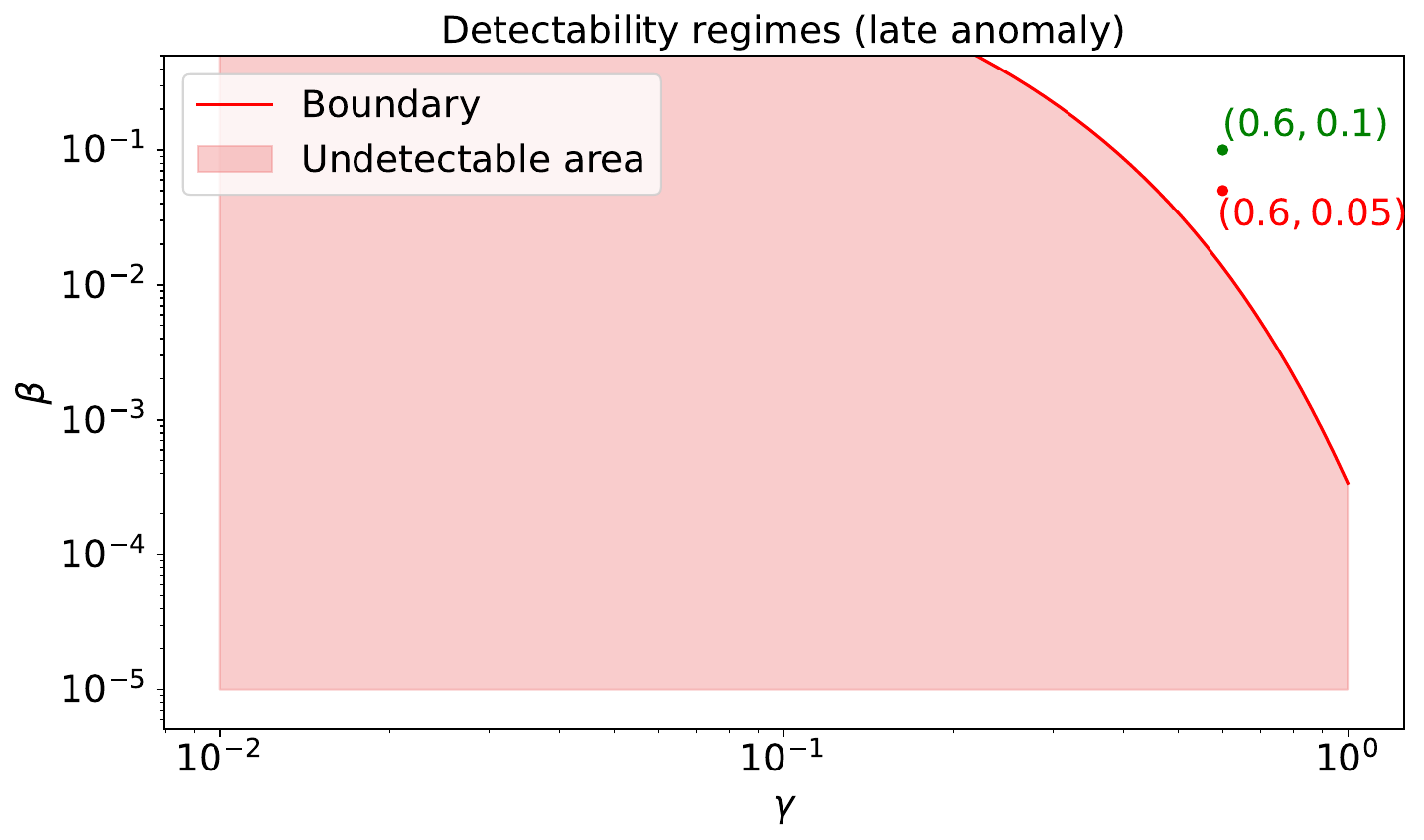}
        \subcaption{$\tau = t - t^{\gamma}$, $t = 10^{4}$}
    \end{minipage}
    \hfill
    \begin{minipage}{0.48\textwidth}
        \centering
        \includegraphics[width=\linewidth]{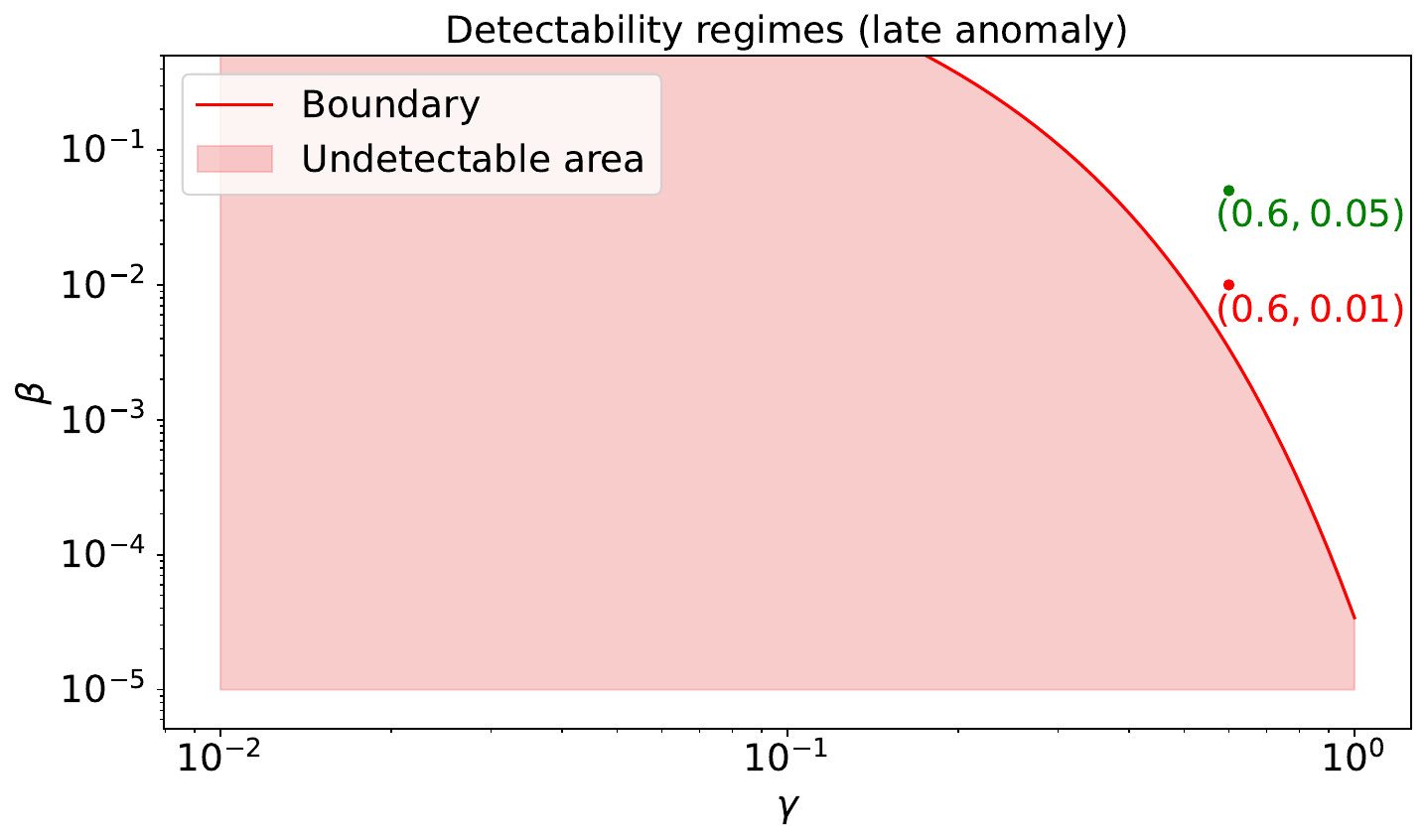}
        \subcaption{$\tau = t - t^{\gamma}$, $t = 10^{5}$}
    \end{minipage}

        \caption{Detectability regimes across different parameter settings. Other parameters are $ m = 3$ and $\delta = 0.1$. In the shaded area, the inequality (18) is violated; the anomaly cannot be detected. In the white area, the inequality (18) is satisfied; this does not guarantee the detection. Green points indicate parameter settings from Table~\ref{tab:Detection_results_known_delta}, for which the detection power is about $80\%$ or higher. Red points indicate parameter settings for which the detection power is substantially below this level.}
    \label{fig:undetectable_region}
\end{figure}

More specifically, the anomalous vertex $v_\tau$ receives an extra edge at each time step with probability at least $\frac{\beta}{2m + \beta + \delta}$.
Consequently, the probability that $v_\tau$ does not receive any additional edge over the time interval $[\tau+1, t]$ is at most
\begin{align}
\label{Prob: no extra edge}
\left( 1 - \frac{\beta}{2m + \beta + \delta} \right)^{m(t-\tau)}.
\end{align}
 When this probability is large, the network evolution becomes difficult to distinguish from that under the null model, making successful detection unlikely. We therefore use 0.2 as a threshold, corresponding to a desired power of at least 0.8:
\begin{align}
\left( 1 - \frac{\beta}{2m + \beta + \delta} \right)^{m(t-\tau)} \leq 0.2.
\end{align}
Taking logarithms on both sides leads to
\begin{align}
m(t-\tau)\log\!\left( 1 - \frac{\beta}{2m + \beta + \delta} \right) \leq \log(0.2).
\end{align}
If this inequality is violated, then the probability of generating no anomalous edges is sufficiently large so that the detection power cannot exceeds 0.8.
Figure~\ref{fig:undetectable_region} illustrates how the model parameters influence the detectable and undetectable regions. Several patterns emerge:
\begin{enumerate}[label=(\arabic*)]
\item The anomaly becomes increasingly difficult to detect as $\beta$ becomes smaller.
\item Detection is challenging when $\tau$ occurs near the end of the network evolution.
\item Increasing the network size from $ t = 10^{4}$ to $ t = 10^{5}$ further enlarges the detectable regions, indicating improved power in larger networks.
\end{enumerate}

\section{Empirical analysis}

In this section, we demonstrate the performance of the iteration estimation procedure and evaluate the effectiveness of the proposed anomaly-detection method. To this end, we first present the estimation results when the anomaly enters the network in different regimes, and then conduct the experiments on detection under various conditions.

\subsection{Candidate screening strategy}

Both parameter estimation and anomaly detection require likelihood maximization across all candidate values ($ \tau = 2,3, \cdots, t$), therefore the number of calls to the most computationally intensive procedure scaling linearly in the network size and becomes prohibitive for large networks. To ensure scalability, we introduce a screening strategy that restricts attention to a data-driven subset of candidate vertices, substantially reducing the computational burden while maintaining detection performance.

A key empirical observation motivating our approach is that an anomalous vertex typically exhibits accelerated degree growth and consequently attains a substantially larger degree than ordinary vertices. Motivated by this structural behavior, we introduce a degree-based screening criterion to identify a reduced set of candidate vertices. Naturally, we select $K$ candidate vertices with highest degrees. However, when the anomaly occurs at a late stage of the network evolution, the anomalous vertex may not yet have accumulated a sufficiently large degree. One way to address this, is to augment the candidate set by including the vertices with top $K$ normalized degree ratio defined by
\begin{align}
\label{eq:ratio}
    \frac{D_i(t) + \delta }{\mathbb{E}_{PA(\delta)}\!\left[D_i(t) + \delta \right]},
\end{align}
where $D_i(t)$ denotes the observed degree of $v_i$ at time $t$, and the expectation is taken under the standard PA model as given in~\eqref{rule-1}. This ratio is motivated by the model-based expected degree and therefore depends on $\delta$. However, candidate selection is used as a preliminary screening step, and we do not want to assume that $\delta$ is known at this stage. To keep this step independent of parameter information, we set $\delta=0$ in this ratio when constructing candidate sets. Specifically, instead of using \eqref{eq:ratio}, we include candidate vertices with top $K$ normalized degree ratio
\begin{align}
\label{eq:ratio0}
    r_i = \frac{D_i(t) }{\mathbb{E}_{PA(0)}\!\left[D_i(t)\right]}.
\end{align}
Since the maximal degree grows quickly, ranking by the ratio \eqref{eq:ratio0} is as effective in catching the anomaly as \eqref{eq:ratio}; this is also confirmed in our preliminary experiments. This rule is used throughout the simulation study, including experiments with known $\delta$.

\begin{figure}[!htbp]
    \centering
    \begin{minipage}{0.48\textwidth}
        \centering
        \includegraphics[width=\linewidth]{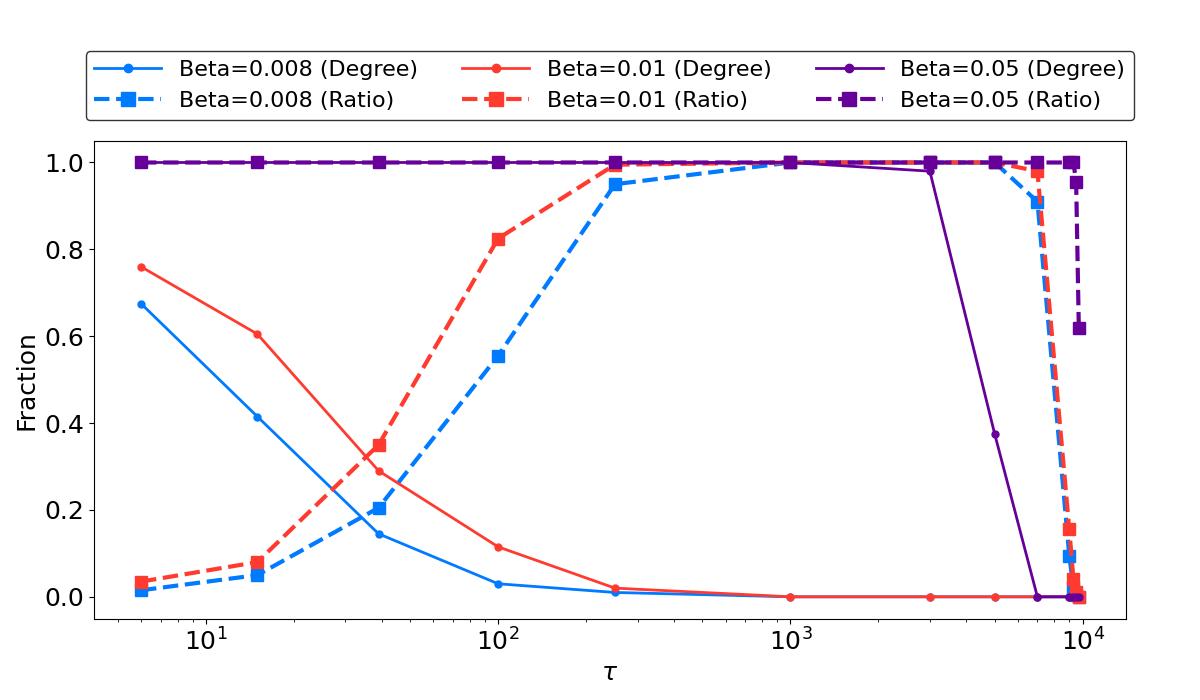}
        \subcaption{Network size: $t = 10^{4}$}
        \label{fig: proportion_top5_small_size}
    \end{minipage}
    \begin{minipage}{0.48\textwidth}
        \centering
        \includegraphics[width=\linewidth]{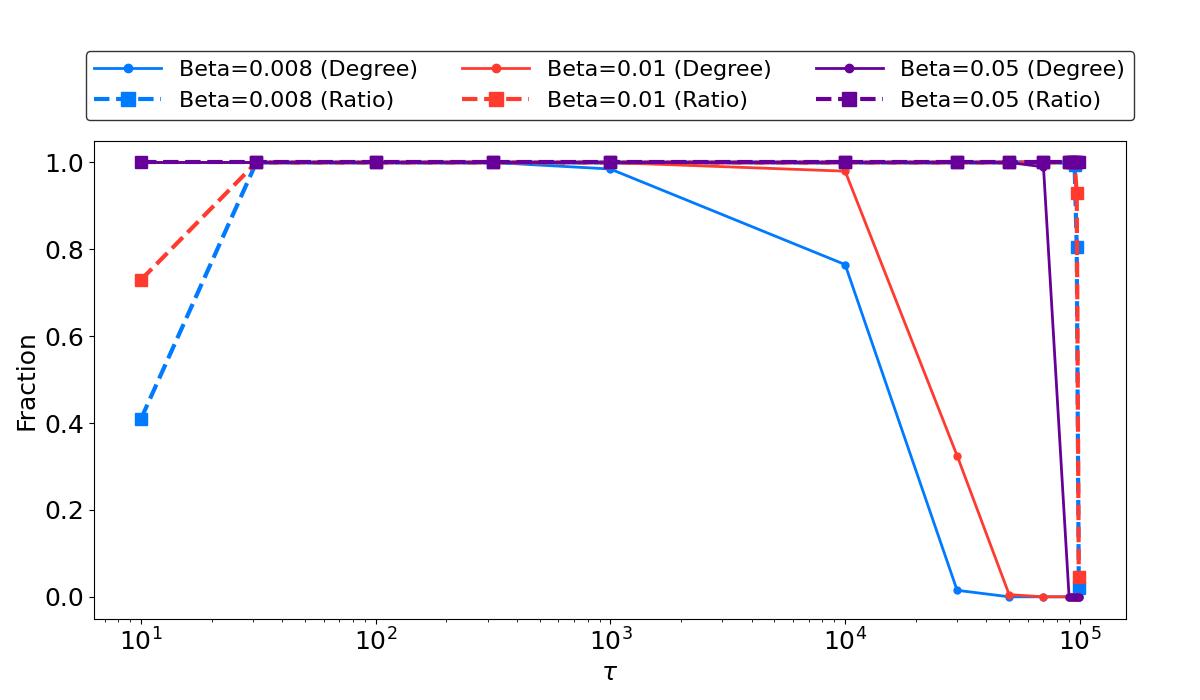}
        \subcaption{Network size: $t = 10^{5}$}
        \label{fig: proportion_top5_big_size}
    \end{minipage}
    \caption{The fraction of simulated graphs in which the anomaly  $v_\tau$ is included among the top-five vertices ranked by degrees (solid lines) or by the ratio with $\delta=0$ (dashed lines). For each parameter pair $(\beta, \tau)$, the fraction is computed over 200 independent replicates. The graphs are generated with $m=3$ and $\delta = 0.1$. The x-axis is shown on a logarithmic scale for $\tau$.}
    \label{fig:Proportion_top5}
\end{figure}

We assess the adequacy of this choice empirically by evaluating the frequency with which the true anomaly is included within the screened candidate set. As observed from the results in Figure~\ref{fig:Proportion_top5}, the anomaly $v_\tau$ is captured with high frequency by either the top five vertices by degree, or the top five vertices by ratio $r_i$. Based on these observations, in the subsequent experiments we consider a common candidate set consisting of the union of the top five candidates ranked by degree and the top five candidates ranked by ratio.

\subsection{Computational implementation}

The simulation study was implemented in Python using custom code based on the model definition. For each parameter setting, an evolving graph was generated by adding one vertex at a time and sampling the endpoints of its $m$ edges according to the specified attachment probabilities under $H_0$ or $H_a$. The intermediate graphs obtained during this construction form the graph sequence used for likelihood evaluation, parameter estimation, and test statistic computation. Numerical operations and optimizations were carried out using standard scientific computing tools, including NumPy and SciPy, with performance-critical simulation routines accelerated using Numba when necessary. The simulation code is available on Github~\cite{code_repo}.

\subsection{Performance of parameters estimation} 

To evaluate the performance of Algorithm~\ref{alg:joint-estimation}, we conduct a set of simulation experiments. We generate networks with size $t = 10^{5}$ vertices, where each arriving vertex forms $m = 3$ edges. The true model parameters are set to $\beta = 0.5$ and $\delta = 0.1$.

As explained in Section~\ref{sec:model}, we assess the estimator under three scenarios: early anomaly ($\tau = \lfloor t^{\gamma} \rfloor$ with $\gamma = 0.4$), midway anomaly ($\tau = \lfloor \gamma t \rfloor $ with $\gamma = 0.5$), and late anomaly ($\tau = \lfloor t - t^{\gamma}\rfloor$ with $\gamma = 0.6$). The estimation results for these settings are reported below.

\subsubsection{Performance of individual parameter estimation}

We first evaluate the estimation performance of each parameter separately. For each parameter of interest, we fix the remaining parameters at their true values and estimate the target parameter via maximum likelihood.

As reported in the Table~\ref{tab:Single_parameter_estimation_performance}, the anomaly location $ \tau$ is identified with $100\%$ accuracy in all cases, when $\beta$ and $\delta$ are given, demonstrating that $\tau$ is well identifiable in this setting. The estimator $\hat{\beta}$ of $\beta$ shows better performance when the anomaly occurs early or midway through the network evolution, as evidenced by smaller standard deviations compared to the late-anomaly case. The behavior is consistent with the cumulative influence of $\beta$ on the attachment dynamics, which becomes more difficult to distinguish the anomaly when it occurs later. Correspondingly, the empirical length of the confidence interval of $\hat{\beta}$ increases. The estimator $\hat{\delta}$ of $\delta$ exhibits a stable pattern: both the standard deviation and the length of empirical confidence interval of $\hat{\delta}$ remain relatively consistent across all scenarios. This stability can be attributed to the fact that $\delta$ influences the entire network evolution, allowing the estimation procedure to exploit information from the fully observed trajectory.

\begin{table}[h]
\centering
\caption{Performance of individual parameter estimation for networks of size $t = 10^{5}$ with $m=3$, based on 300 simulated datasets. For the discrete parameter $\tau$, we report the recovery rate, defined as the proportion of datasets for which $\hat{\tau} = \tau$.}
\label{tab:Single_parameter_estimation_performance}
\begin{tabular}{cccccc}
\toprule
Case & Parameter & True values & Mean & Std & 5\% - 95\% / recovery rate\\
\midrule
\multirow{3}{*}{Early anomaly}
 & $\beta$  & 0.5 & 0.4999 & 0.0046 & [0.4918, 0.5065] \\
 & $\delta$ & 0.1 & 0.0995 & 0.0161 & [[0.0708, 0.1261] \\
 & $\tau$   & 100    & --    & --    & 1.00 \\
\midrule
\multirow{3}{*}{Midway anomaly}
 & $\beta$  & 0.5 & 0.5002 & 0.0054 & [0.4911, 0.5089] \\
 & $\delta$ & 0.1 & 0.0993 & 0.0163 & [0.0729, 0.1260] \\
 & $\tau$   & \num{50000}    & --    & --    & 1.00 \\
\midrule
\multirow{3}{*}{Late anomaly}
 & $\beta$  & 0.5 & 0.5016 & 0.0329 & [0.4463, 0.5556] \\
 & $\delta$ & 0.1 & 0.0982 & 0.0181 & [0.0678, 0.1273] \\
 & $\tau$   & \num{99000}    & --    & --    & 1.00 \\
\bottomrule
\end{tabular}
\end{table}

\begin{figure}[!htbp]
    \centering
    % Early anomaly (tau = 100) ---
    \begin{minipage}{0.42\textwidth}
        \centering
        \includegraphics[width=\linewidth]{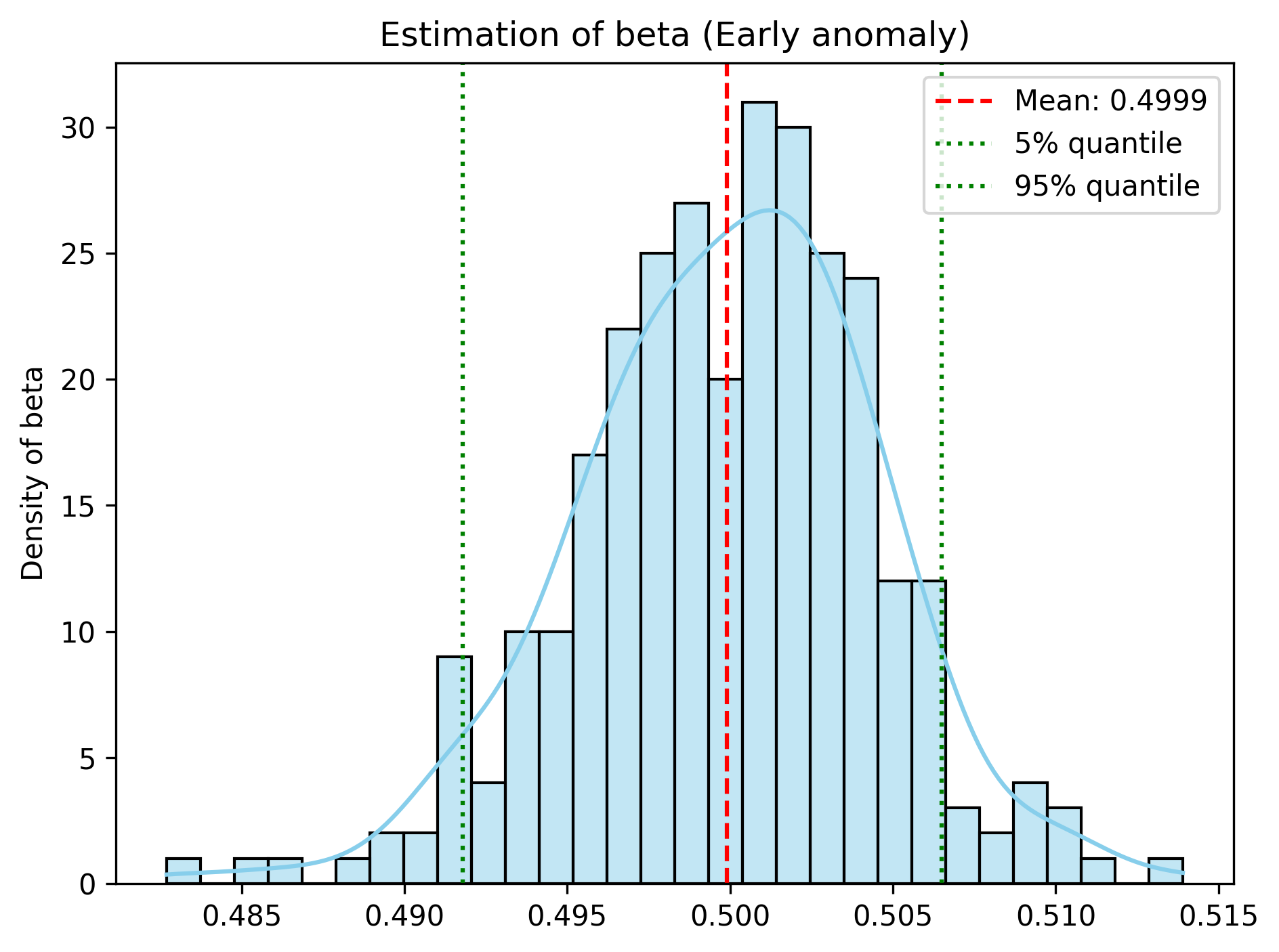}
        \subcaption{MLE of $\beta$ with $\delta$ fixed ($\beta = 0.5$).}
        \label{fig:Early_anomaly_beta_estimate_1}
    \end{minipage}
    \hspace{0.5cm}
    \begin{minipage}{0.42\textwidth}
        \centering
        \includegraphics[width=\linewidth]{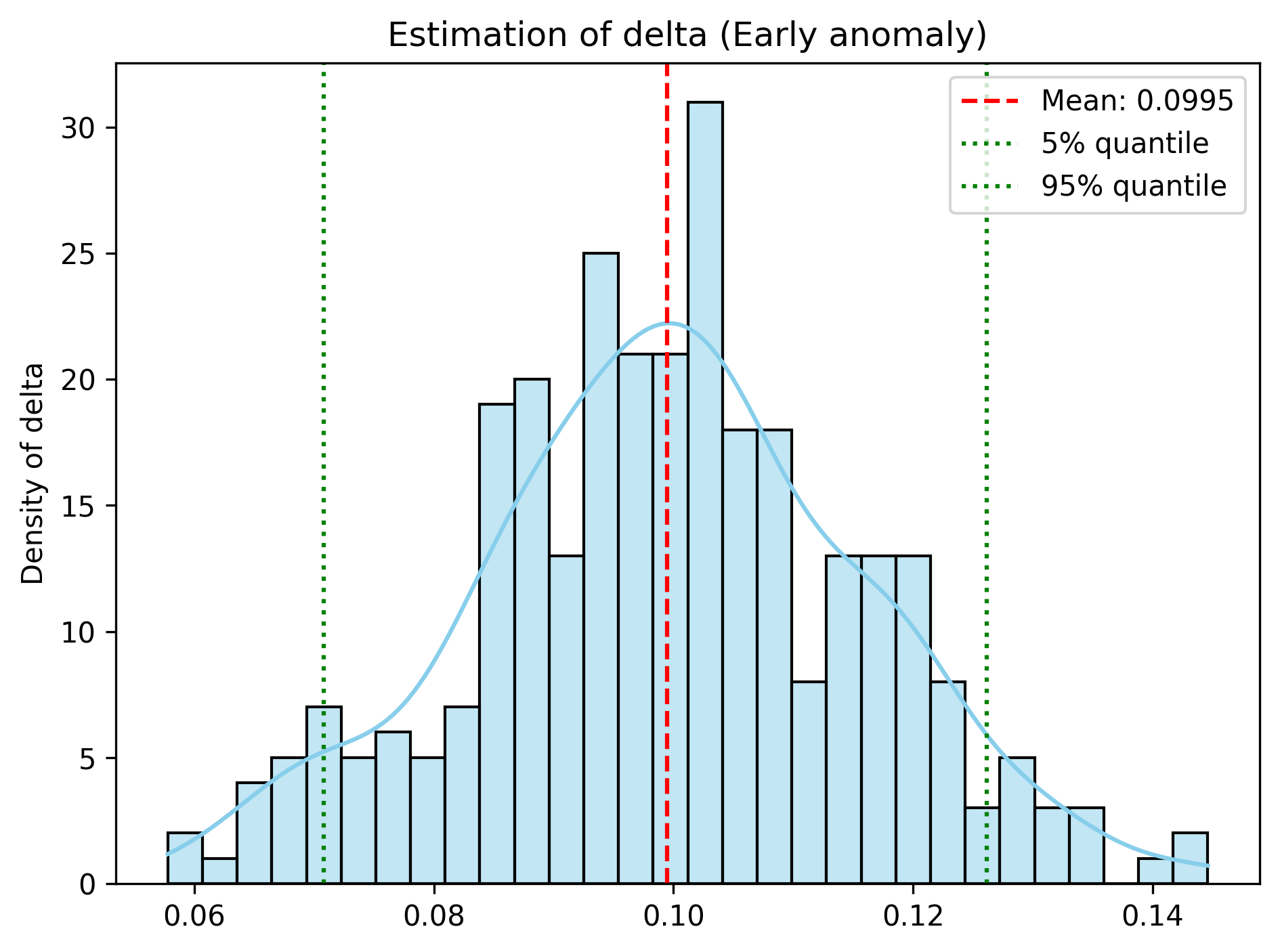}
        \subcaption{MLE of $\delta$ with $\beta$ fixed ($\delta = 0.1$).}
        \label{fig:Early_anomaly_delta_estimate_1}
    \end{minipage}

    \vspace{0.5cm}

    % mid-way anomaly (tau = 50000) ---
    \begin{minipage}{0.42\textwidth}
        \centering
        \includegraphics[width=\linewidth]{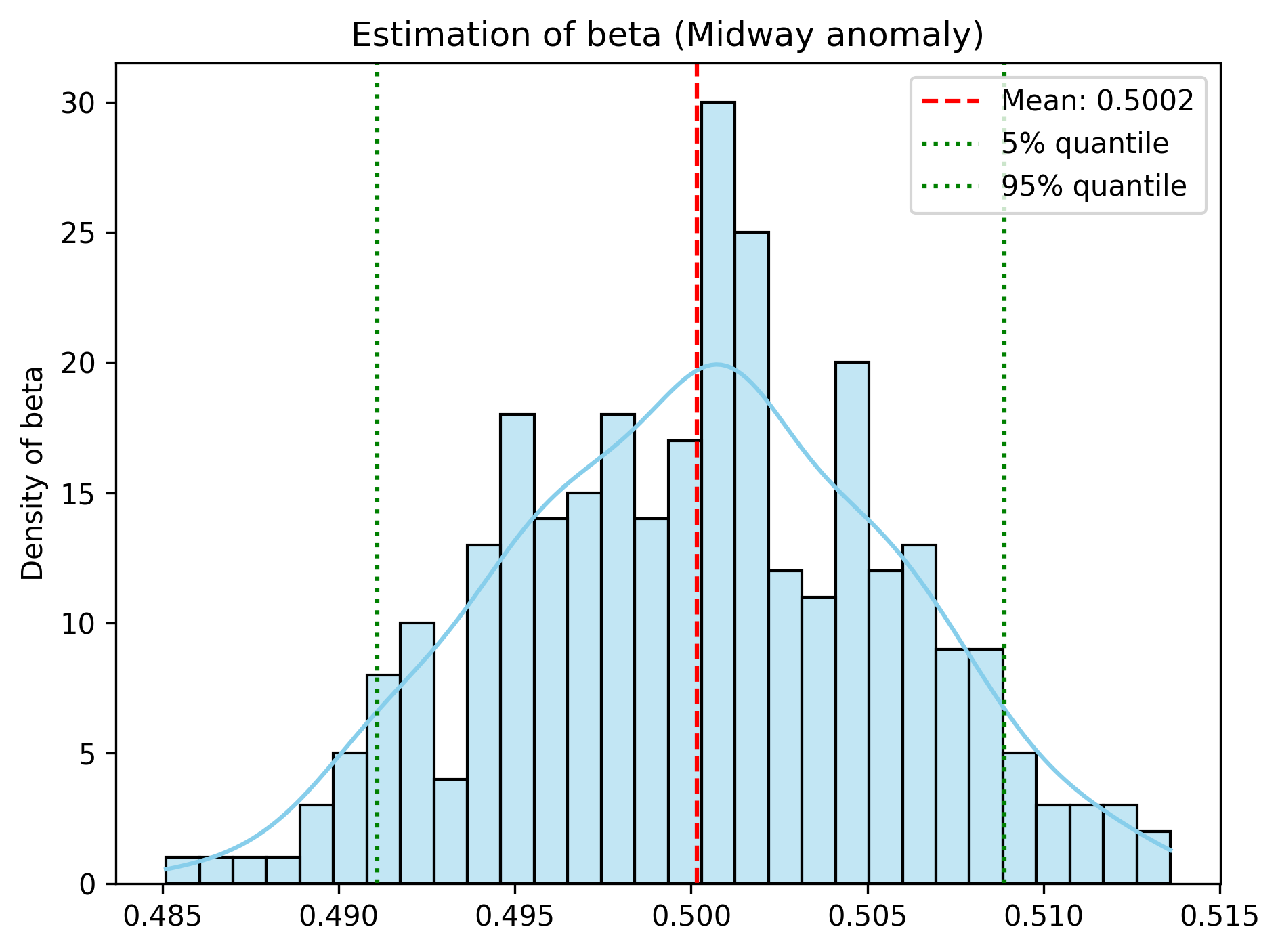}
        \subcaption{MLE of $\beta$ with $\delta$ fixed ($\beta = 0.5$).}
        \label{fig:Midway_anomaly_beta_estimate_1}
    \end{minipage}
    \hspace{0.5cm}
    \begin{minipage}{0.42\textwidth}
        \centering
        \includegraphics[width=\linewidth]{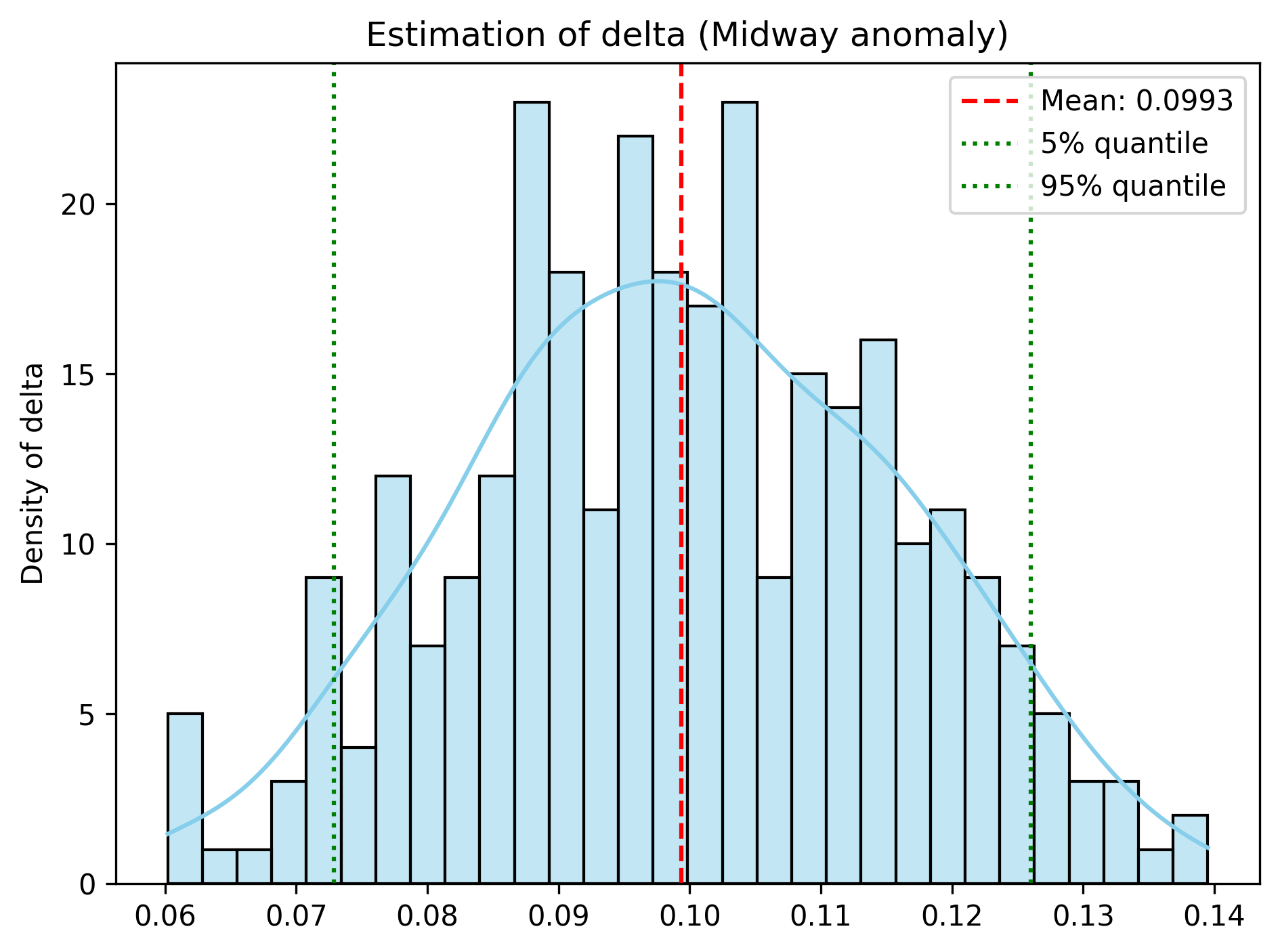}
        \subcaption{MLE of $\delta$ with $\beta$ fixed ($\delta = 0.1$).}
        \label{fig:Midway_anomaly_delta_estimate_1}
    \end{minipage}

    \vspace{0.5cm}

    % late anomaly (tau = 99000) ---
    \begin{minipage}{0.42\textwidth}
        \centering
        \includegraphics[width=\linewidth]{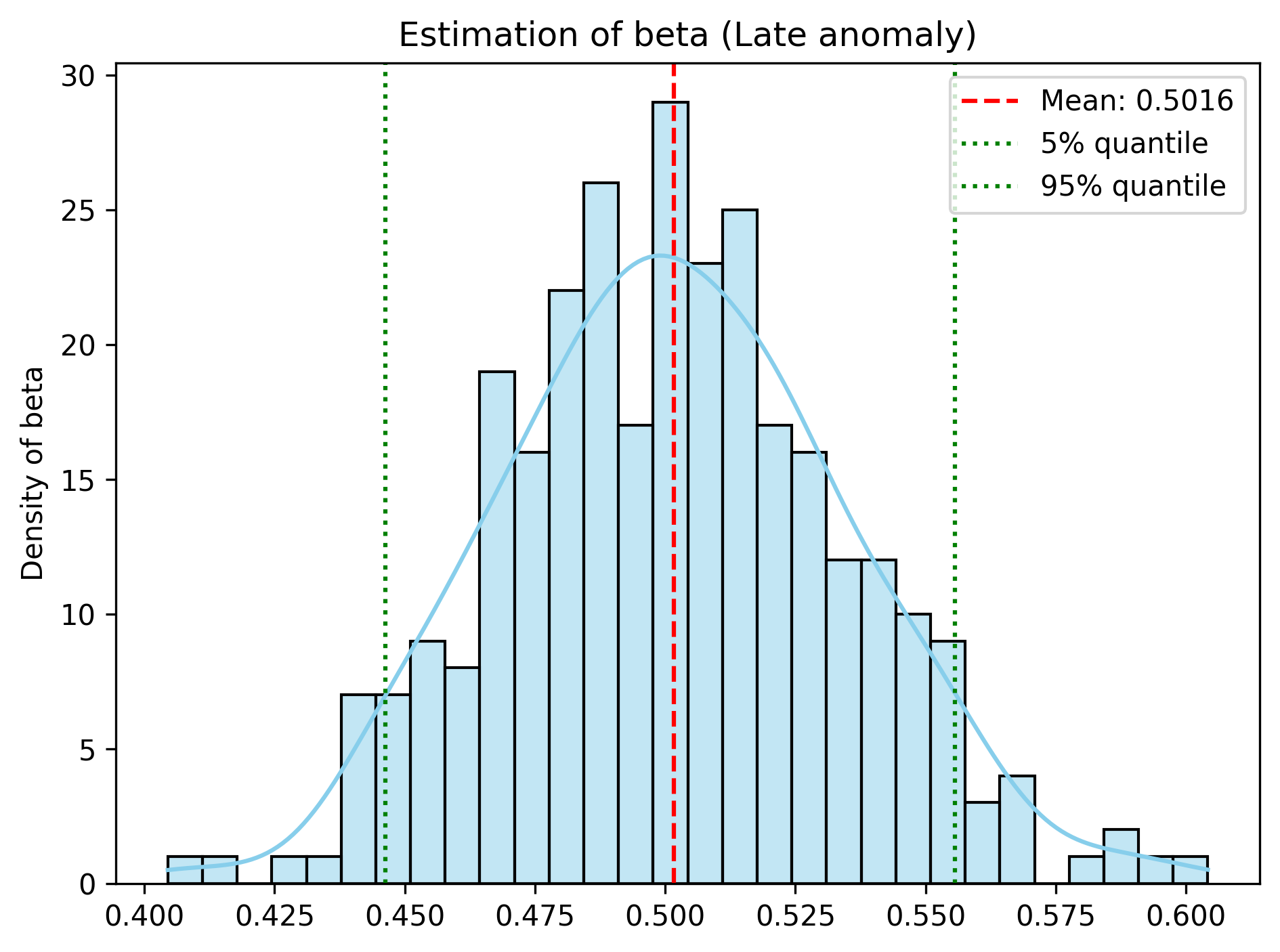}
        \subcaption{MLE of $\beta$ with $\delta$ fixed ($\beta = 0.5$).}
        \label{fig:late_anomaly_beta_estimate_1}
    \end{minipage}
    \hspace{0.5cm}
    \begin{minipage}{0.42\textwidth}
        \centering
        \includegraphics[width=\linewidth]{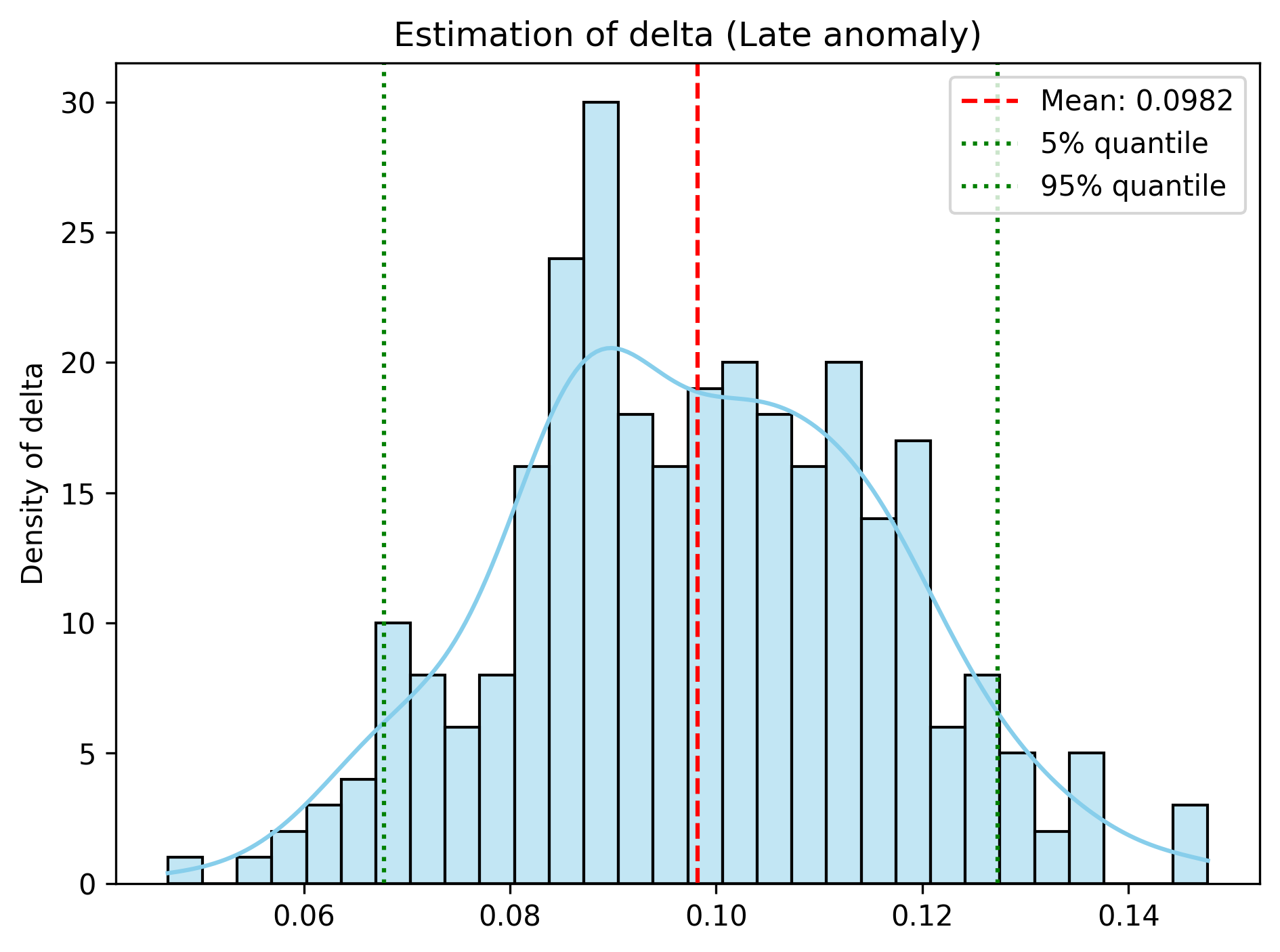}
        \subcaption{MLE of $\delta$ with $\beta$ fixed ($\delta = 0.1$).}
        \label{fig:late_anomaly_delta_estimate_1}
    \end{minipage}

    \caption{Histogram plots of 300 replicates of the MLE obtained from \emph{individual estimation}. Simulation parameters are $t = 10^{5}$, $m=3$, $\beta=0.5$, and $\delta=0.1$. Early anomaly: $\tau = \lfloor t^{\gamma} \rfloor$ with $\gamma = 0.4$; midway anomaly: $\tau = \lfloor \gamma t \rfloor $ with $\gamma = 0.5$; late anomaly: $\tau = \lfloor t - t^{\gamma}\rfloor$ with $\gamma = 0.6$.}
    \label{fig:MLE_conditional_estimation}
\end{figure}

Figures~\ref{fig:MLE_conditional_estimation} shows the empirical distributions of the estimator of $\beta$ and $\delta$ across the three cases. In all scenarios, the distributions are centered close to the true parameter value when the remaining parameters are known.

\subsubsection{Performance of joint estimation}

\label{subsec: Joint estimation performance}
We next investigate the performance of the proposed estimation procedure, where for each candidate vertex, $\beta$ and $\delta$ are estimated jointly via maximum likelihood, and the anomaly arrival time $\tau$ is identified as the candidate vertex that yields the highest maximized likelihood.

In our implementation, $\beta$ is optimized under the constraint that it lies in $(10^{-6}, 10)$. This upper bound is chosen as a numerical range that is substantially larger than the parameter values considered in the experiments (e.g., $\beta = 0.5$). We set the tolerance rate as $10^{-5}$ and employ five distinct initial values for $\beta$, given by $(1.0, 3.0, 5.0, 7.0, 9.0)$ spanning the interval $(10^{-6}, 10)$ to reduce sensitivity to initialization and improve coverage of the parameter space.

\begin{figure}[!htbp]
    \centering
    % Early anomaly
    \begin{minipage}{0.45\textwidth}
        \centering
        \includegraphics[width=\linewidth]{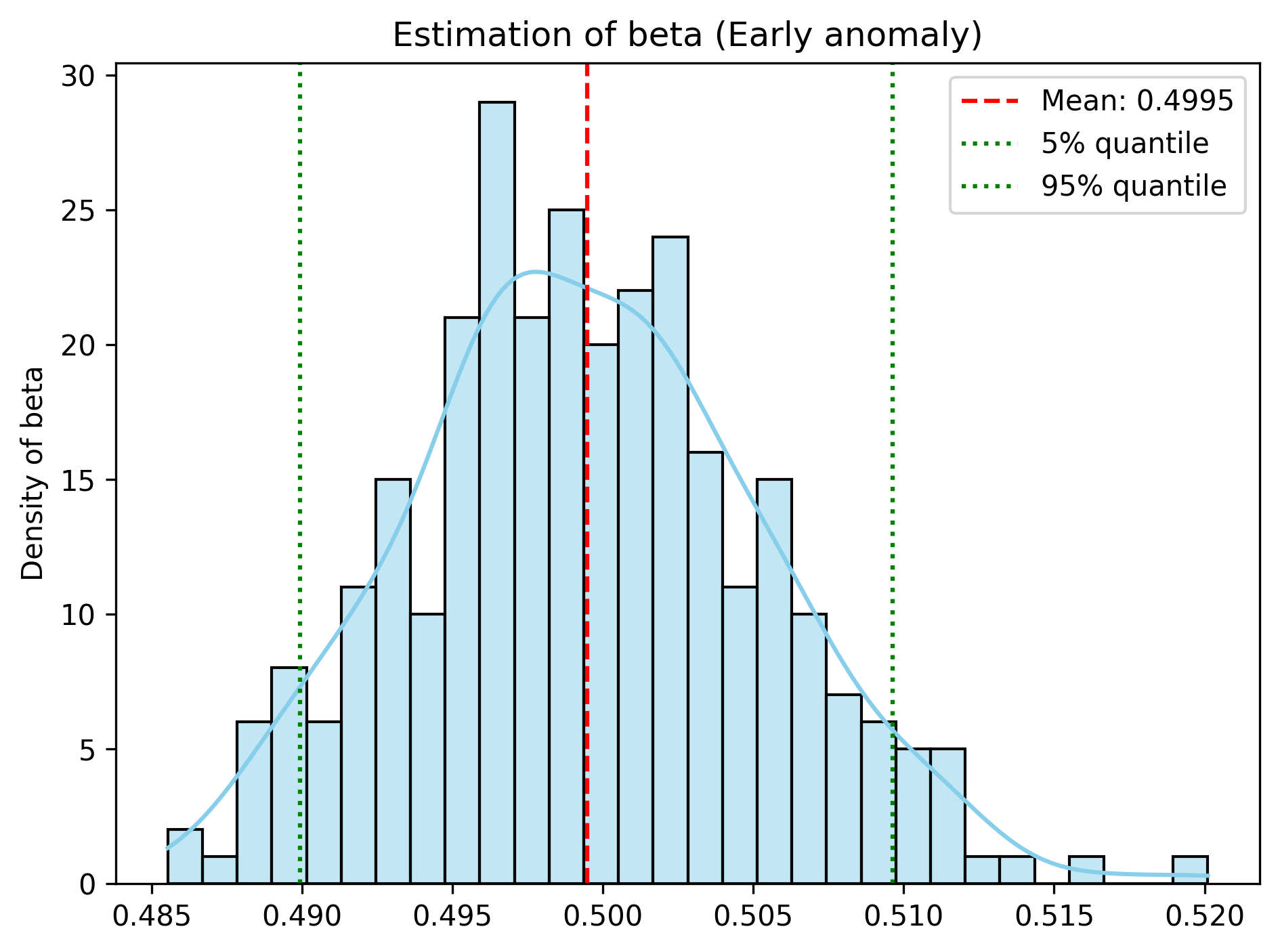}
        \subcaption{Joint estimate of $\beta$ ($\beta = 0.5$).}
        \label{fig:Early_anomaly_beta_estimate_2}
    \end{minipage}
    \hspace{0.5cm}
    \begin{minipage}{0.45\textwidth}
        \centering
        \includegraphics[width=\linewidth]{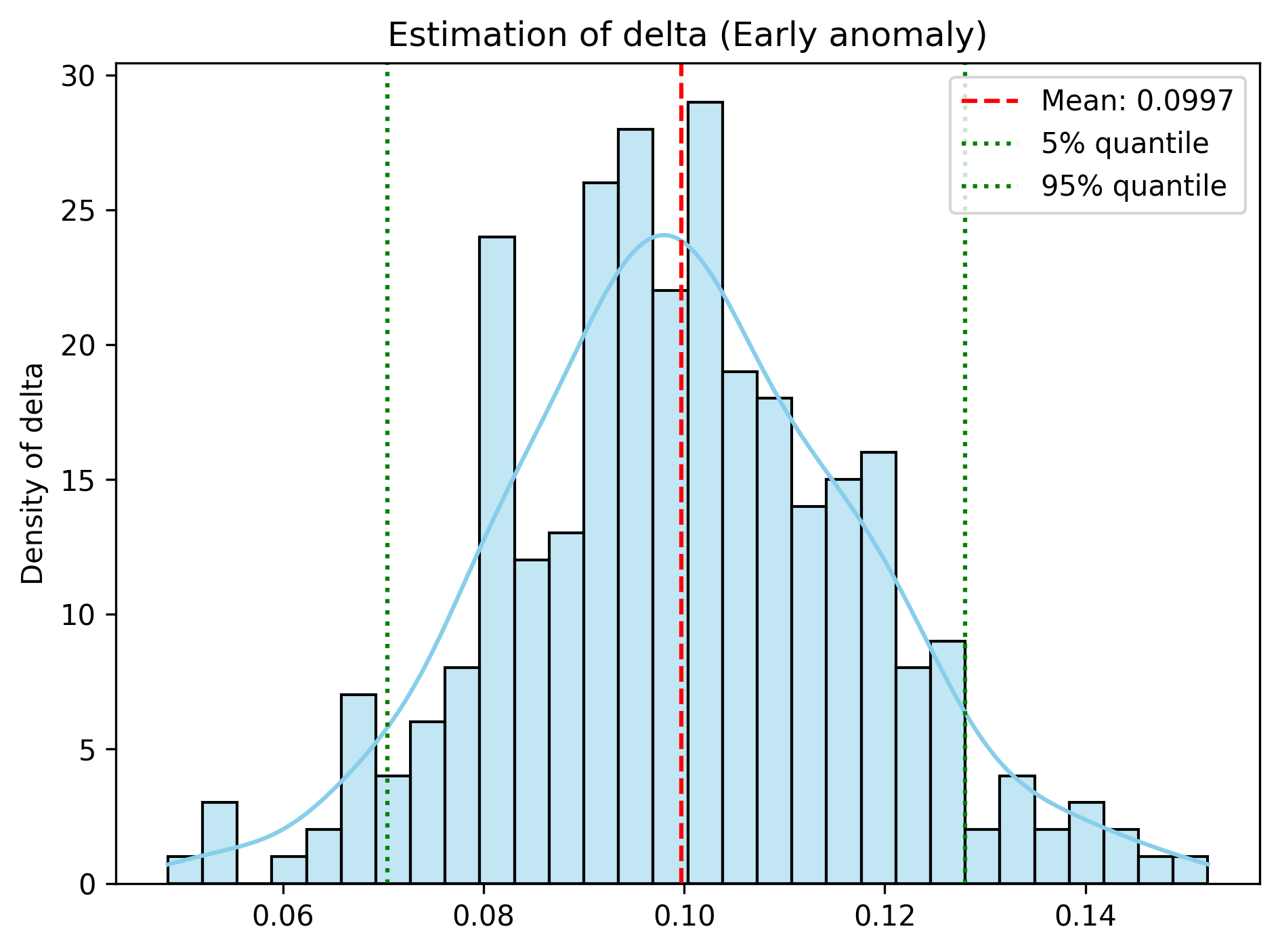}
        \subcaption{Joint estimate of $\delta$ ($\delta = 0.1$).}
        \label{fig:Early_anomaly_delta_estimate_2}
    \end{minipage}

    \vspace{0.5cm} 

    % mid-way anomaly
    \begin{minipage}{0.45\textwidth}
        \centering
        \includegraphics[width=\linewidth]{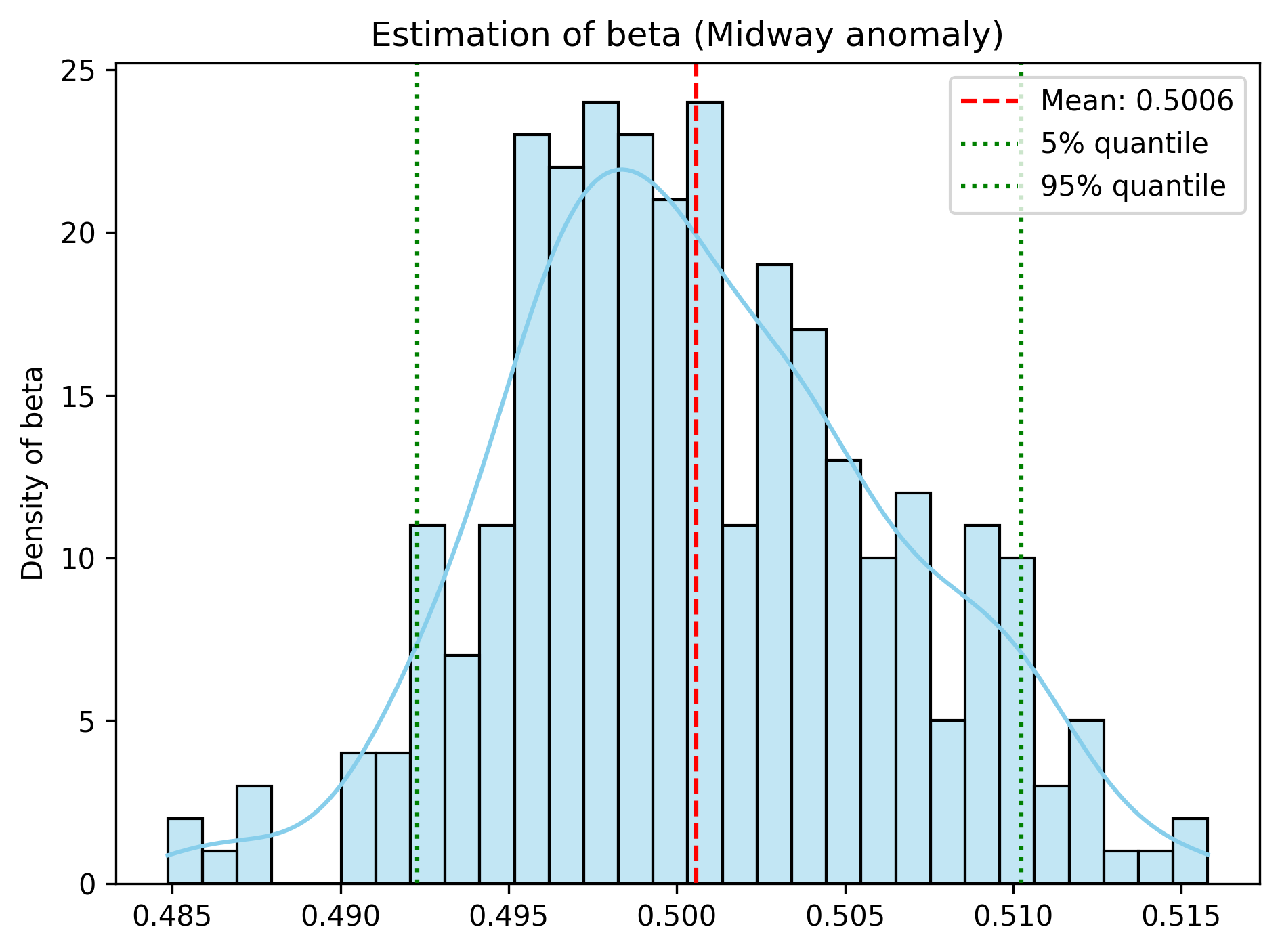}
        \subcaption{Joint estimate of $\beta$ ($\beta = 0.5$).}
        \label{fig:Midway_anomaly_beta_estimate_2}
    \end{minipage}
    \hspace{0.5cm}
    \begin{minipage}{0.45\textwidth}
        \centering
        \includegraphics[width=\linewidth]{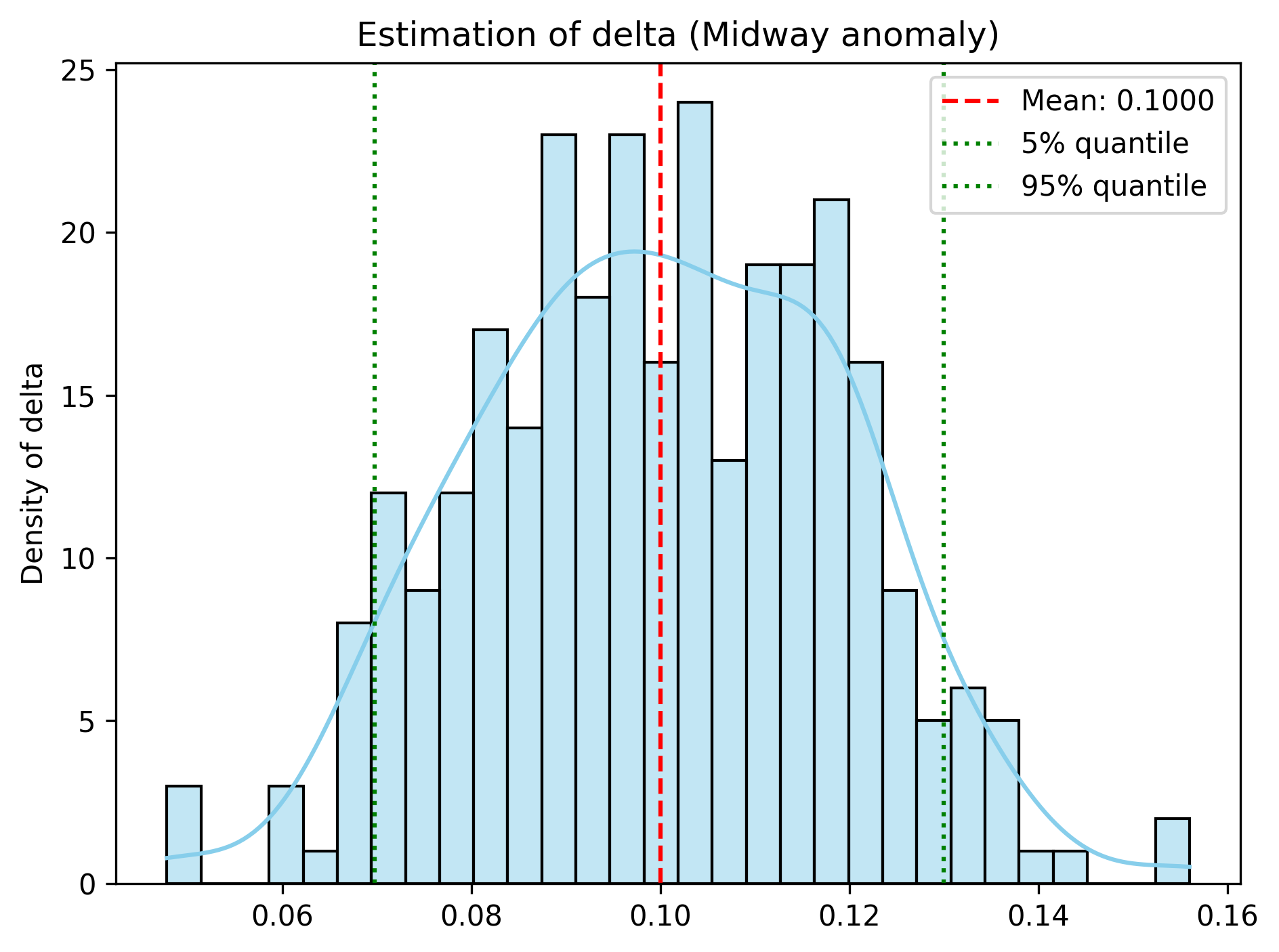}
        \subcaption{Joint estimate of $\delta$ ($\delta = 0.1$).}
        \label{fig:Midway_anomaly_delta_estimate_2}
    \end{minipage}

    \vspace{0.5cm}

    % late anomaly
    \begin{minipage}{0.45\textwidth}
        \centering
        \includegraphics[width=\linewidth]{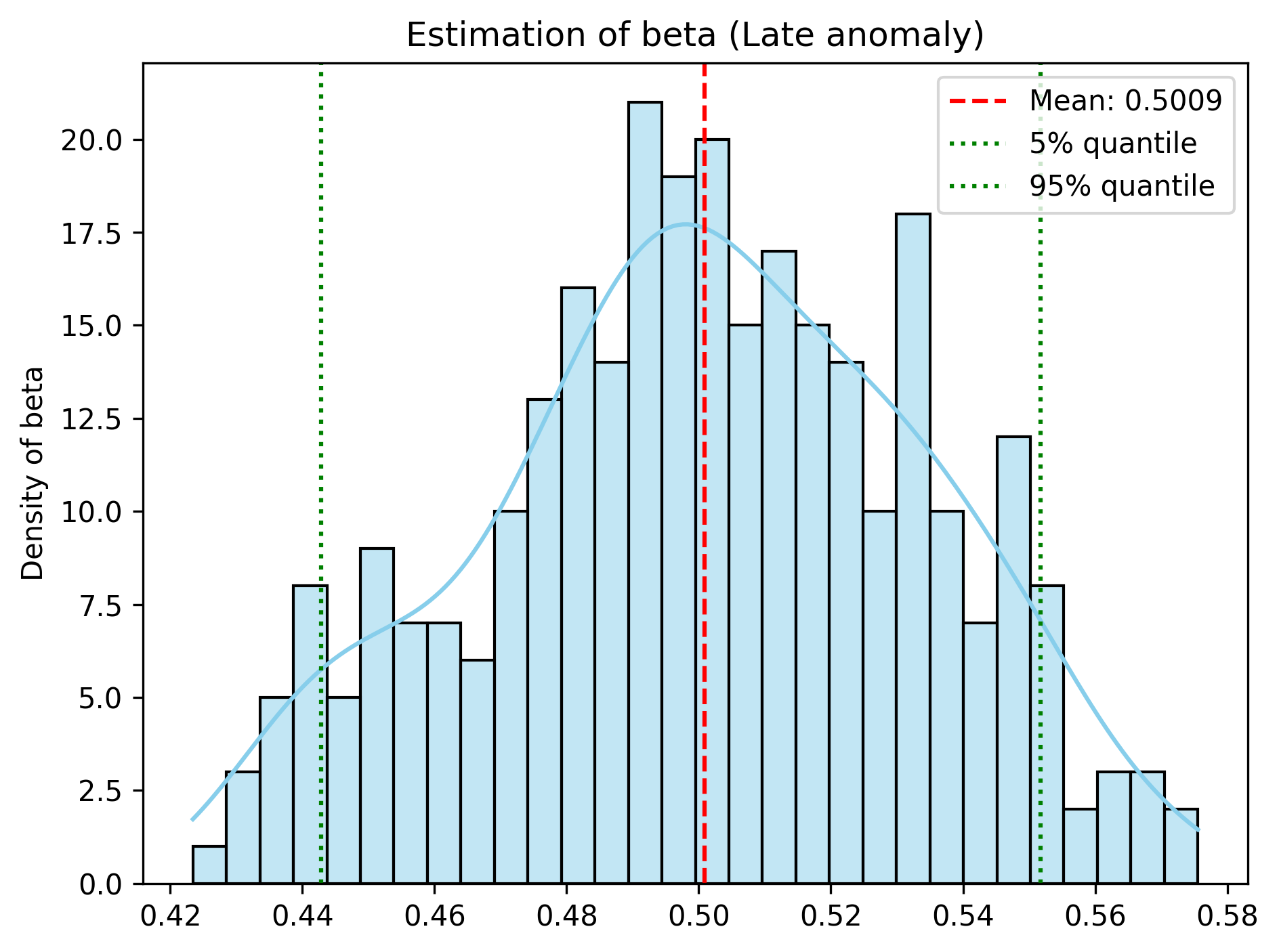}
        \subcaption{Joint estimate of $\beta$ ($\beta = 0.5$).}
        \label{fig:late_anomaly_beta_estimate_2}
    \end{minipage}
    \hspace{0.5cm}
    \begin{minipage}{0.45\textwidth}
        \centering
        \includegraphics[width=\linewidth]{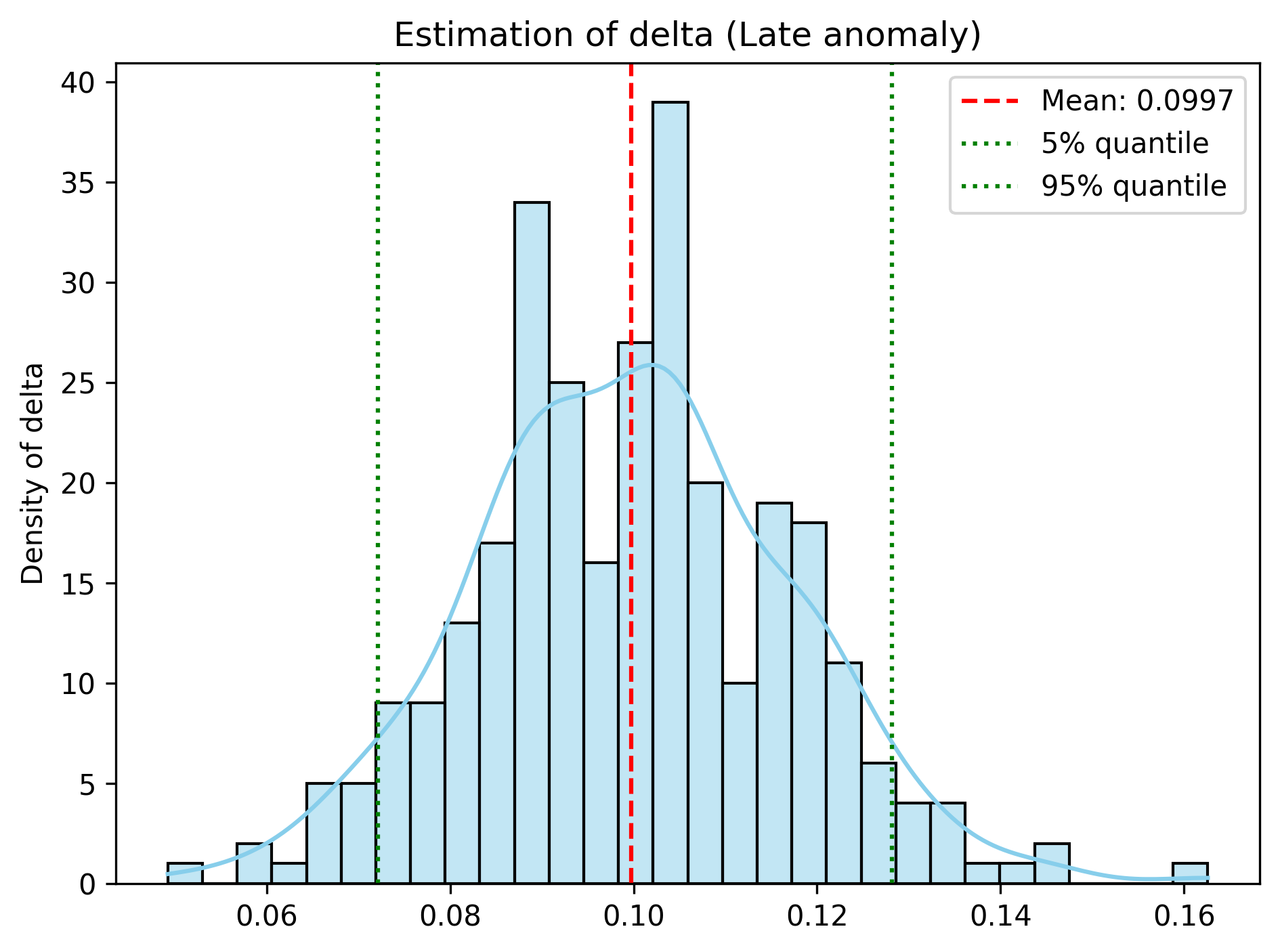}
        \subcaption{Joint estimate of $\delta$ ($\delta = 0.1$).}
        \label{fig:late_anomaly_delta_estimate_2}
    \end{minipage}

    \caption{Histogram plots of 300 replicates of the MLE obtained from \emph{joint estimation}. Simulation parameters are $t = 10^{5}$, $m=3$, and $\beta=0.5, \delta=0.1$. Early anomaly: $\tau = \lfloor t^{\gamma} \rfloor$ with $\gamma = 0.4$; midway anomaly: $\tau = \lfloor \gamma t \rfloor $ with $\gamma = 0.5$; late anomaly: $\tau = \lfloor t - t^{\gamma}\rfloor$ with $\gamma = 0.6$.}
    \label{fig:MLE_joint_estimation}
\end{figure}

Figures~\ref{fig:MLE_joint_estimation} presents the empirical distributions of the estimators of $\beta$ and $\delta$ obtained from joint estimation. The estimates remain centered near the true parameter value across all three cases, indicating that joint estimation does not introduce substantial bias. In particular, the bias, standard deviation and the interval length are comparable to those observed under individual parameter estimation; the results are shown in Table~\ref{tab:Joint_estimation_performance}.

In addition, the anomaly arrival time $\tau$ is identified with 100\% accuracy rate in all cases. For the estimator of $\beta$, the standard deviation and the empirical width of the confidence interval increase in late anomaly scenarios; however, the changes remain moderate and do not compromise estimation stability. The reason for this behavior is that the anomaly affects only a short terminal window, providing less information to estimate the parameter $\beta$.

\begin{table}[htbp]
\centering
\caption{Performance of joint estimation for networks of size $t = 10^{5}$ with $m=3$, based on 300 simulated datasets. For the discrete parameter $\tau$, we report the recovery rate, defined as the proportion of datasets for which $\hat{\tau} = \tau$.}
\label{tab:Joint_estimation_performance}
\begin{tabular}{cccccc}
\toprule
Case & Parameter & True values & Mean & Std & 5\%--95\% / recovery rate \\
\midrule
\multirow{3}{*}{Early anomaly}
 & $\beta$  & 0.5 & 0.4995 & 0.0058 & [0.4899,0.5096 ] \\
 & $\delta$ & 0.1 & 0.0997 & 0.0177 & [0.0704,0.1280] \\
 & $\tau$   & 100 & -- & -- & 1.00 \\
\midrule
\multirow{3}{*}{Midway anomaly}
 & $\beta$  & 0.5 & 0.5006 & 0.0057 & [0.4923, 0.5102] \\
 & $\delta$ & 0.1 & 0.1000 & 0.0192 & [0.0697, 0.1300] \\
 & $\tau$   & \num{50000} & -- & -- & 1.00 \\
\midrule
\multirow{3}{*}{Late anomaly}
 & $\beta$  & 0.5 & 0.5009 & 0.0328  &  [0.4428,0.5517] \\
 & $\delta$ & 0.1 & 0.0997 & 0.0171 & [0.0721, 0.1282] \\
 & $\tau$   & \num{99000} & -- & -- & 1.00 \\
\bottomrule
\end{tabular}
\end{table}
%\FloatBarrier

\begin{figure}[!htbp]
    \centering
    % Early anomaly
    \begin{minipage}{0.48\textwidth}
        \centering
        \includegraphics[width=\linewidth]{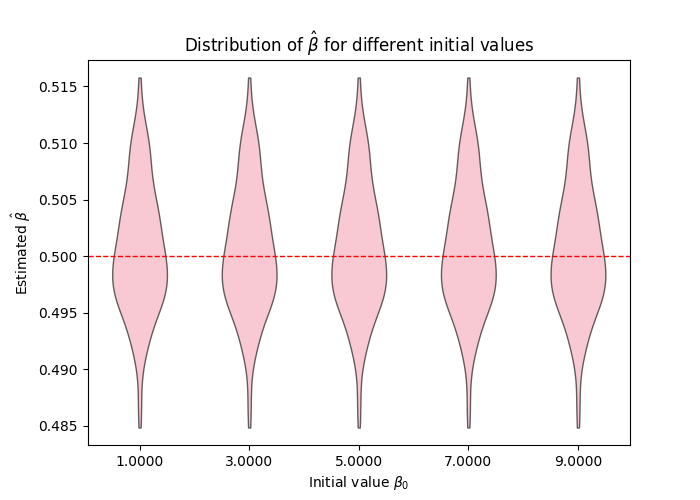}
        \subcaption{$\beta = 0.5$}
        \label{fig:violin_beta}
    \end{minipage}
    %\hspace{0.5cm}
    \begin{minipage}{0.48\textwidth}
        \centering
        \includegraphics[width=\linewidth]{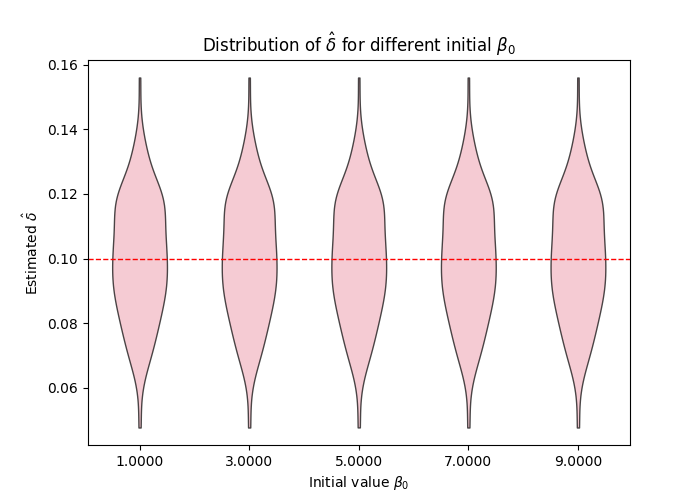}
        \subcaption{$\delta = 0.1$}
        \label{fig:violin_delta}
    \end{minipage}
    \caption{Estimation of 300 replicates obtained from \emph{joint estimation} using different initial values of $\beta$. Simulation parameters are $t = 10^{5}$, $\tau = \num{50000}$, $m = 3$. Violin plots (pink shapes) showing the empirical distribution of the estimations, with the shape's width reflecting the frequency of the estimated values.}

    \label{fig:comparison_different_init_beta}
\end{figure}

However, using multiple initial values for $\beta$ to estimate the parameters is computationally expensive, as the optimization must be repeated for each initial value over the same network. Therefore, we analyze the estimator distributions across different initial values of $\beta$. Figure~\ref{fig:comparison_different_init_beta} shows the empirical distributions of $\hat{\beta}$ and $\hat{\delta}$ for the midway anomaly case based on different initial values of $\beta$, and similar patterns are observed in the other cases. The distributions are consistent across all initializations, indicating that the joint procedure is relatively insensitive to the choice of initial values for $\beta$. Accordingly, we select an initial value of $\beta$ close to the midpoint of the interval (specifically, $\beta_0 = 5.0$) in the detection experiment with unknown $\delta$ for computational efficiency.

Overall, the empirical results indicate that joint estimation remains sufficiently accurate for use in the subsequent detection procedure.

\subsection{Performance of anomaly detection}

We evaluate the proposed detection method using synthetic networks generated according to the PA mechanism. We consider both the setting in which the parameter $\delta$ is known and a more realistic scenario in which $\delta$ must be estimated. 
Detection and localization performance are assessed using two empirical measures. The empirical power is defined as the proportion of simulated datasets generated under the alternative hypothesis for which the null hypothesis is rejected. The localization rate is defined as the proportion of replicates in which the estimated anomaly location $\hat{\tau}$ exactly matches the true value of $\tau$.

\subsubsection{Detection performance with known \texorpdfstring{$\delta$}{delta}}

We first evaluate the detection performance under the setting where $\delta$ is known. Specifically, we examine the empirical power of the proposed test across different values of $\beta$ and under various anomaly scenarios. 
Throughout this experiment, we fix $\delta = 0.1$ and $m = 3$, and consider the two network sizes, $t = 10^{4}$ and $t = 10^{5}$, to assess the effect of network size on detection performance. We consider three representative values of $\beta$, corresponding to weak, moderate, and strong anomalous attachment effects.
For each configuration, we generate 300 independent replicates under $H_a$ and evaluate detection and localization performance across different anomaly settings. 
The significance level is fixed at $\alpha = 0.05$, and the critical value for each anomaly scenario is obtained as the empirical $(1-\alpha)$ quantile of the test statistic based on 300 independent replicate networks under $H_0$ with the fixed $\delta$. Since $\delta$ is known, the replicates under $H_0$ are drawn from the correct distribution and therefore the probability of the type-I error converges to $\alpha$ when the number of replicates goes to infinity.

\begin{table}[ht]
\centering
\caption{Performance of anomaly detection with known $\delta$. Bold entries indicate the parameter settings displayed in Figure~\ref{fig:undetectable_region}. Early anomaly: $\tau = \lfloor t^{\gamma} \rfloor$ with $\gamma = 0.2$; midway anomaly: $\tau = \lfloor \gamma t \rfloor $ with $\gamma = 0.5$; late anomaly: $\tau = \lfloor t - t^{\gamma}\rfloor$ with $\gamma = 0.6$.}
\label{tab:Detection_results_known_delta}
\begin{tabular}{c|c|ccc|ccc}
\toprule
& & \multicolumn{3}{c|}{$t = 10^4$} & \multicolumn{3}{c}{$t = 10^5$} \\
\cmidrule(lr){3-5} \cmidrule(lr){6-8}
Case & $\beta$ & $\tau$ & Power & Localization & $\tau$ & Power & Localization \\
\midrule
\multirow{6}{*}{Early anomaly}
 & 0.001 & 6  & 0.0667  & 0.00    & \textbf{10} & \textbf{0.0667} & \textbf{0.0267} \\
 & 0.005 & 6  & 0.0733  & 0.0400  & \textbf{10} & \textbf{0.9867} & \textbf{0.9833}\\
 & 0.01  & \textbf{6}  & \textbf{0.3600}  & \textbf{0.5133}    & 10 & 1.0    & 1.0 \\
 & 0.05  & \textbf{6}  & \textbf{1.0}     & \textbf{1.0}     & 10 & 1.0    & 1.0\\
 & 0.1   & 6  & 1.0     & 1.0     & 10 & 1.0    & 1.0\\
 & 0.5   & 6  & 1.0     & 1.0     & 10 & 1.0    & 1.0\\
\midrule
\multirow{7}{*}{Midway anomaly}
& 0.0001 & 5000 & 0.0733 & 0.00 & \textbf{\num[detect-weight=true]{50000}} & \textbf{0.0567} & \textbf{0.0033}\\
 & 0.001 & \textbf{5000}  & \textbf{0.0733} & \textbf{0.0133}  & \textbf{\num[detect-weight=true]{50000}} & \textbf{0.9900} & \textbf{0.9933} \\
 & 0.005 & \textbf{5000}  & \textbf{0.7700}   & \textbf{0.8333} & \num{50000} & 1.0    & 1.0 \\
 & 0.01  & 5000  & 1.0 & 1.0   & \num{50000} & 1.0    & 1.0 \\
 & 0.05  & 5000  & 1.0    & 1.0    & \num{50000} & 1.0    & 1.0\\
 & 0.1   & 5000  & 1.0    & 1.0    & \num{50000} & 1.0    & 1.0 \\
 & 0.5   & 5000  & 1.0    & 1.0    & \num{50000} & 1.0    & 1.0\\
\midrule
\multirow{6}{*}{Late anomaly}
 & 0.01 & 9748   & 0.0567 & 0.00   & \textbf{\num[detect-weight=true]{99000}} & \textbf{0.1167} & \textbf{0.0600} \\
 & 0.05 & \textbf{9748}   & \textbf{0.5233}   & \textbf{0.4800} & \textbf{\num[detect-weight=true]{99000}} & \textbf{1.0}    & \textbf{1.0} \\
 & 0.1  & \textbf{9748}   & \textbf{0.9733}  & \textbf{0.9733}   & \num{99000} & 1.0    & 1.0\\
 & 0.5  & 9748   & 1.0    & 1.0    & \num{99000} & 1.0    & 1.0\\
 & 0.7  & 9748   & 1.0    & 1.0    & \num{99000} & 1.0    & 1.0\\
 & 0.9  & 9748   & 1.0    & 1.0    & \num{99000} & 1.0    & 1.0\\
\bottomrule
\end{tabular}
\end{table}

The resulting detection performance is summarized in Table~\ref{tab:Detection_results_known_delta}. We see that detection and localization performance depend on both the anomaly arrival time $\tau$ and the anomaly strength $\beta$. Among the three regimes, the proposed test exhibits reduced empirical power and localization accuracy when $\beta$ is small. This is consistent with the shape of the detectability regime. However, we also notice that the anomaly can remain difficult to detect even outside of the undetectable region.  For larger values of $\beta$,  both power of the test and localization rate achieve 100\% accuracy  reflecting the stronger influence of anomalous attachment on the degree growth of the anomalous vertex. 

We also observe that the localization rate is not always consistent with the empirical power. This is due to the two-stage nature of the detection procedure: we first screen for candidate vertices based on screening strategy, and then compare the likelihood ratio against a critical value. As a result, it is possible that the test correctly rejects $H_0$ (high empirical power) but fails to identify the correct vertex (low localization rate), or vice versa.

\subsubsection{Detection performance with unknown \texorpdfstring{$\delta$}{delta}}
\label{subsec: Detection unknown delta}
We next consider a setting in which the parameter $\delta$ is unknown and must be estimated from the data. The parameters are the same as in the experiments where $\delta$ is known, with $\delta = 0.1$, $m = 3$, and $\alpha = 0.05$, except that we restrict to the smaller network size $ t = 10^{4}$ due to computational reasons. We first validate that the test is well calibrated under $H_0$ through the empirical type-I error simulation procedure. We generate $N_1 = 300$ network replicates under $H_0$ with $t = 10^{4}$, and for each we generate $100$ PA networks under $H_0$ using $\hat{\delta}_1$ estimated under $H_a$ to approximate the critical value. The empirical type-I error is approximately 5.67\%, close to the nominal level of 5\%.

The presented empirical analysis of the detection procedure is computationally intensive. To illustrate the computational cost, consider a single setting with $ \beta = 0.01$, $\tau = 100$ and $t = 10^{4}$. In order to estimate the power of the test, we run the entire procedure  $100$ times for each choice of $\beta$ and $\tau$. Specifically, we generate 100 PA network replicates under $H_a$ using rule~\eqref{rule-2}. For each replicate, we estimate $(\hat{\beta}_1, \hat{\delta}_1, \hat{\tau}_1)$ under $H_a$ and estimate $\hat{\delta}_0$ under $H_0$, and then generate 100 null networks using $\hat{\delta}_1$ to approximate the critical value. This results in a total of $100 \times 100 = \num{10000}$ network generations and a corresponding number of numerical optimizations per setting, which is computationally demanding, particularly for large networks.

To reduce the computational cost, we initialize $\beta$ close to the midpoint of the search interval $(10^{-6}, 10)$, $\beta_0 = 5.0$, rather than using multiple initial values. This choice is supported by our earlier finding in Section~\ref{subsec: Joint estimation performance} that the joint estimation procedure is insensitive to the choice of initial value. Additionally, we adapt the candidate vertex set based on the specific setting:
\begin{enumerate}
    \item[(1)] For early anomaly case, we define the candidate set as the union of two top-five lists, one ranked by degree and the other by the ratio $r_i$. Figure~\ref{fig: proportion_top5_small_size} shows that, when $\beta$ is small, $v_\tau$ appears less frequently in the ratio-based top five but more frequently in the degree-based top five over time; however, neither fraction exceeds 0.8. This suggests that neither criterion alone is reliable with smaller $\beta$. In contrast, when $\beta$ is sufficiently large, the ratio-based top-five vertices contains the anomaly. Therefore, we use both criteria for the early anomaly case.

    \item[(2)] For the midway and late anomaly case, we instead consider the five vertices with the largest ratio $r_i$ as candidates to reduce the computational burden of estimating the power of the test. This choice is based on the observation in Figure~\ref{fig:Proportion_top5}, when the anomaly occurs relatively late, the anomalous vertex is almost always included among the top five vertices ranked by $r_i$. Nevertheless, in practice application, we recommend considering candidates selected by both criteria, namely the top-ranked vertices by degree and by ratio.
\end{enumerate}

To systematically assess the robustness of the proposed detection procedure, we vary both the anomaly strength parameter $\beta$ and the anomaly arrival time $\tau$, which is controlled through a scaling parameter $\gamma$. The parameter $\gamma$ is chosen differently across anomaly regimes in order to capture the distinct temporal characteristics of early, midway, and late anomalies.

\begin{figure}[htbp]
    \centering
    % --- Early Anomaly ---
    \begin{minipage}{0.45\textwidth}
        \includegraphics[width=\linewidth]{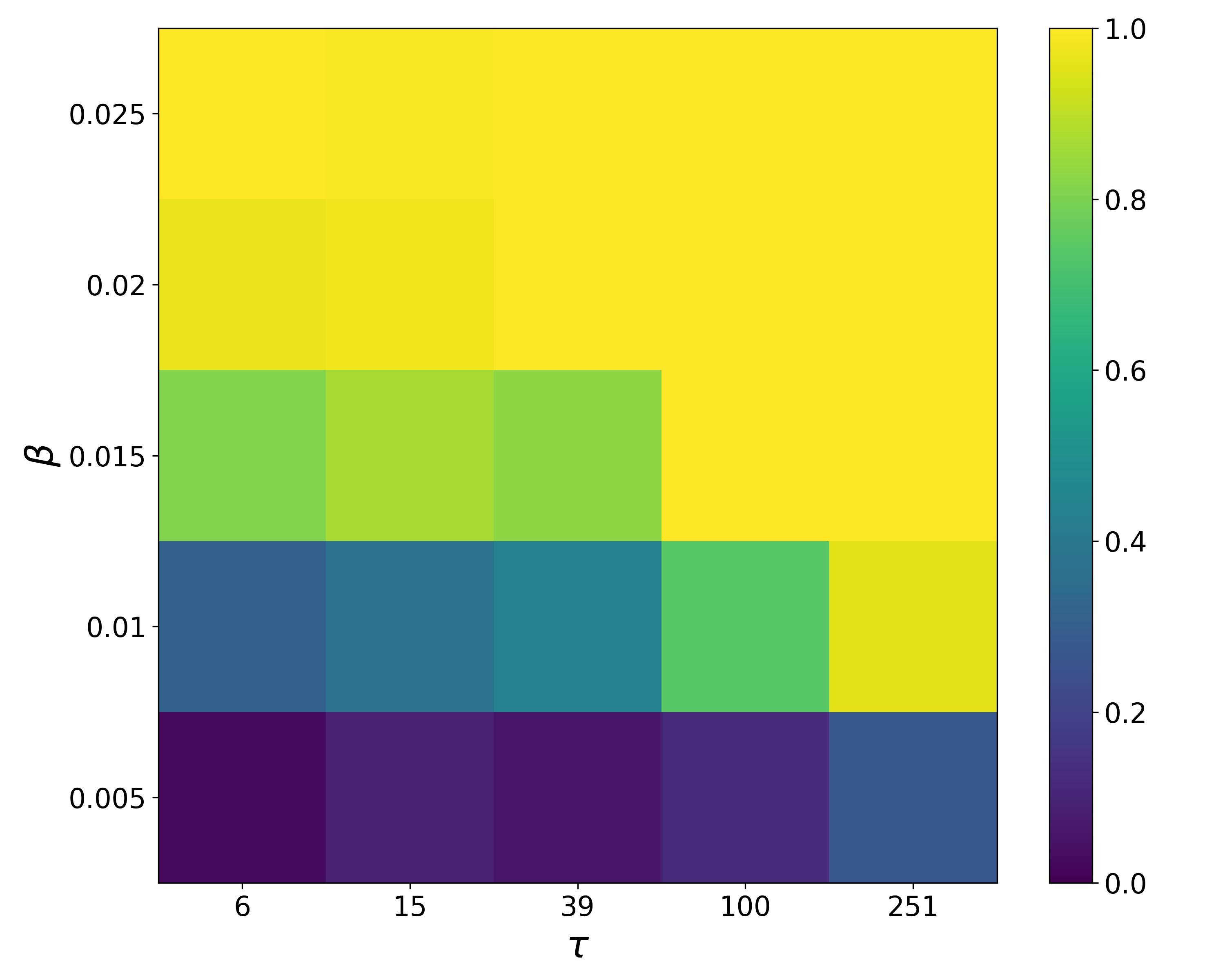}
        \subcaption{Empirical power (Early anomaly)}
    \end{minipage}
    \begin{minipage}{0.45\textwidth}
        \includegraphics[width=\linewidth]{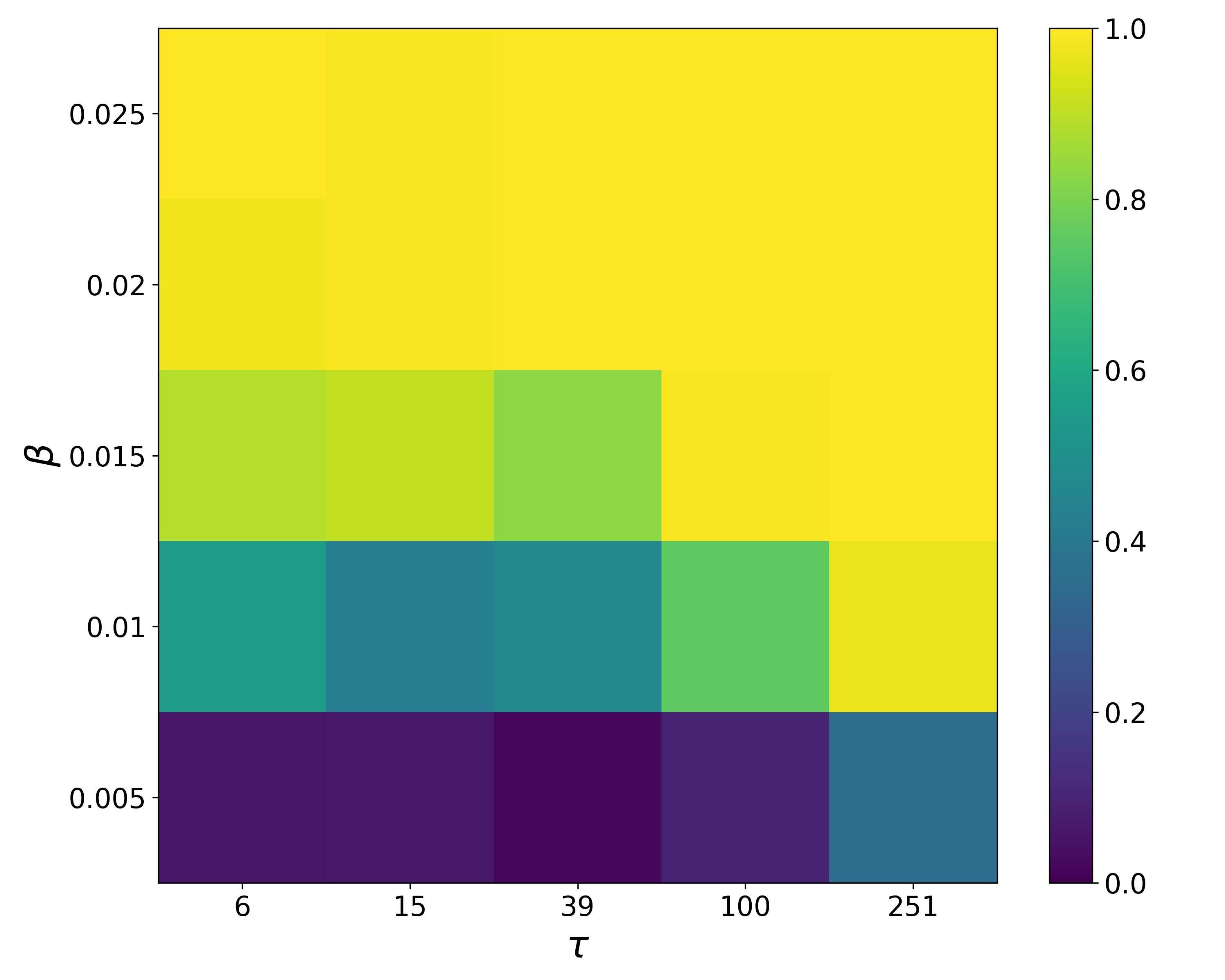}
        \subcaption{Localization rate (Early anomaly)}
    \end{minipage}
    \vspace{0.5cm} 
    
    % --- Mid-way Anomaly ---
    \begin{minipage}{0.45\textwidth}
        \includegraphics[width=\linewidth]{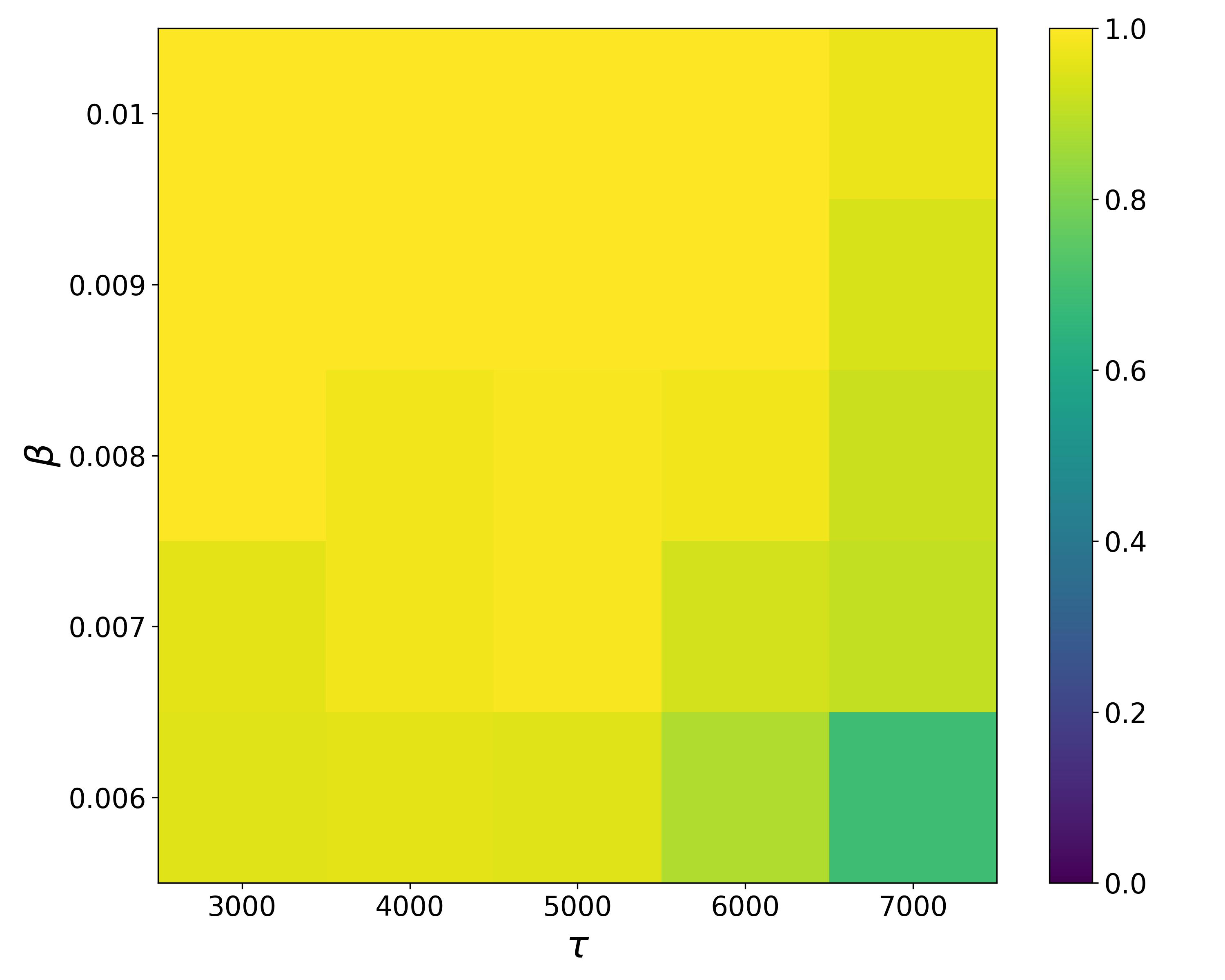}
        \subcaption{Empirical power (Midway anomaly)}
    \end{minipage}
    \begin{minipage}{0.45\textwidth}
        \includegraphics[width=\linewidth]{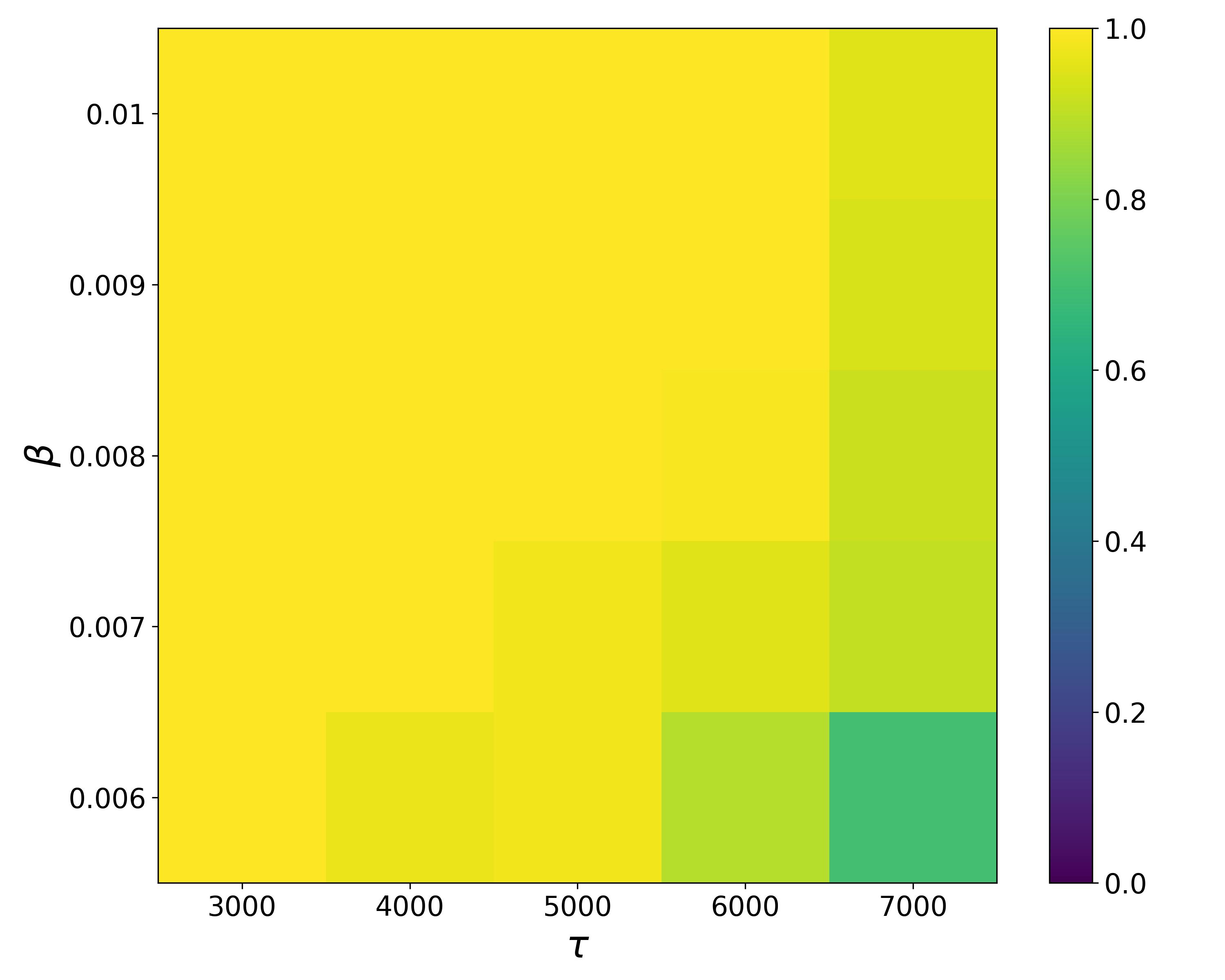}
        \subcaption{Localization rate (Midway anomaly)}
    \end{minipage}
    \vspace{0.5cm} 
    
    % --- Late Anomaly ---
    \begin{minipage}{0.45\textwidth}
        \includegraphics[width=\linewidth]{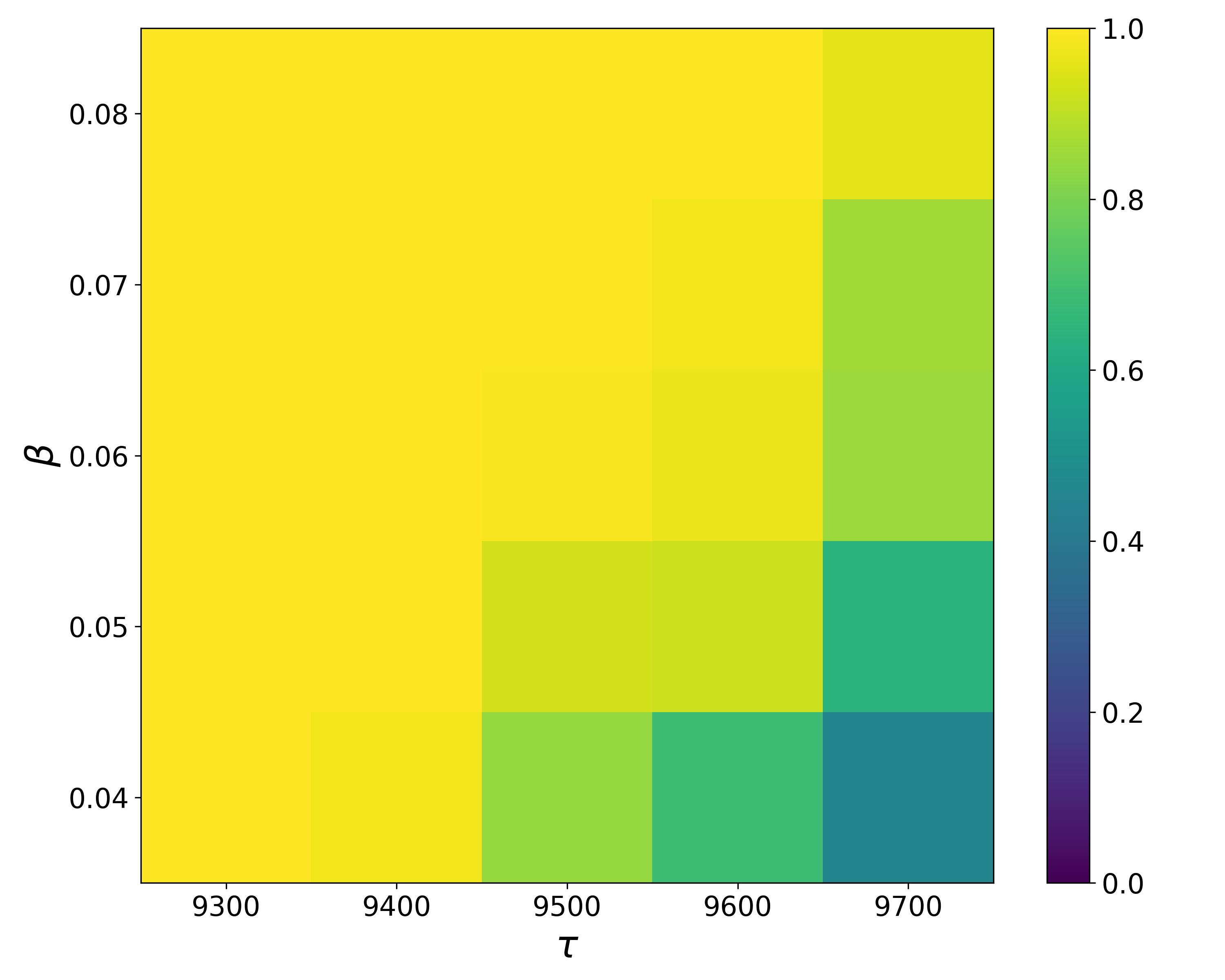}
        \subcaption{Empirical power (Late anomaly)}
    \end{minipage}
    \begin{minipage}{0.45\textwidth}
        \includegraphics[width=\linewidth]{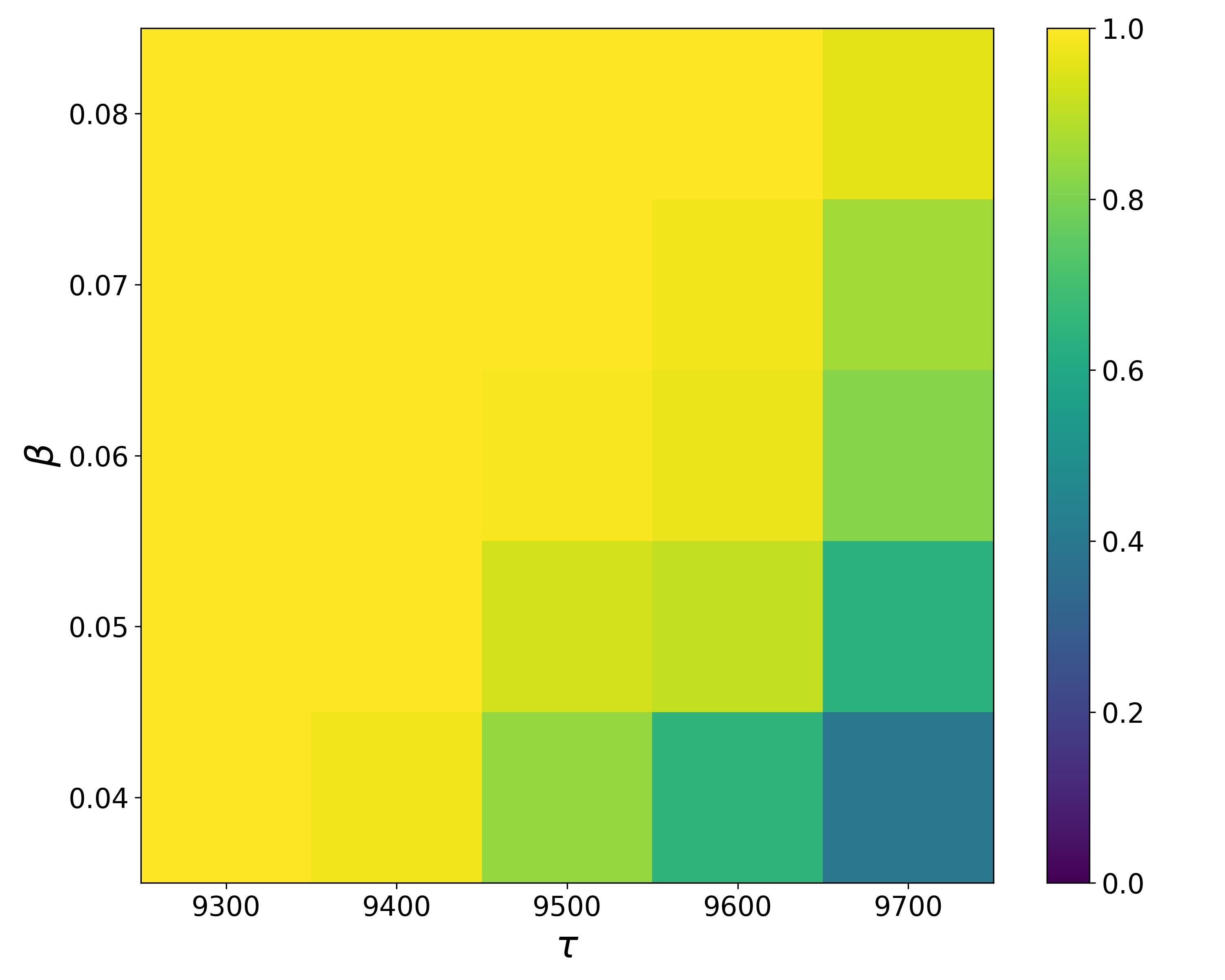}
        \subcaption{Localization rate (Late anomaly)}
    \end{minipage}
    \caption{Detection performance for early, midway, and late anomalies with unknown $\delta$. Parameters are $t = 10^{4}$, $\delta = 0.1$, $m =3$. Early anomaly: $\tau = \lfloor t^{\gamma} \rfloor$, $\gamma \in \{0.2, 0.3, 0.4, 0.5, 0.6\}$; Midway anomaly: $\tau = \lfloor \gamma t \rfloor$, $\gamma \in \{0.3, 0.4, 0.5, 0.6, 0.7\}$; Late anomaly: $\tau = \lfloor t - t^{\gamma} \rfloor$, $\gamma \in \{0.6192, 0.6505, 0.6747, 0.6945, 0.7112\}$.}
    \label{fig:all_heatmaps}
\end{figure}

Figure~\ref{fig:all_heatmaps} shows the detection results for the above cases. We observe several patterns:
\begin{enumerate}[label=(\arabic*)]
    \item Both the empirical power and the localization rate increase monotonically as $\beta$ increases, across all three anomaly regimes, as can be expected.
    \item In the early anomaly regime, detection is particularly challenging when the anomaly occurs at a very early stage and $\beta$ is small, as the anomalous vertex has not yet accumulated sufficient degree to stand out from normal vertices.
    \item In the midway and late anomaly regimes, for a fixed $\beta$, detection becomes more difficult as $\tau$ approaches the end of the network evolution. This is because the anomalous vertex has less time to accumulate excess edges, making its behavior more similar to the baseline model.
\end{enumerate}

The above results demonstrate that the proposed framework performs well even for relatively small values of $\beta$ across different anomaly regimes. We next extend our experiments to a larger network setting with $t = 10^{5}$, employing the same screening strategy, to further assess its ability to detect weak anomalies. Compared to the smaller network setting, detection performance improves substantially, as shown in Table~\ref{tab:Detection_results_big_size}, even for smaller values of $\beta$, across early, midway, and late anomaly regimes. This shows that a larger network size provides more information for distinguishing anomalous attachment behavior, enabling reliable detection even under weaker anomaly strength.

\begin{table}[htbp]
\centering
\caption{Performance of anomaly detection with unknown $\delta$ under $t = 10^{5}, m = 3, \delta = 0.1$.}
\label{tab:Detection_results_big_size}
\begin{tabular}{ccccc}
\toprule
Case & $\tau$ & $\beta$ & Empirical power & Localization rate \\
\midrule
Early anomaly & 10 & 0.005 & 0.97 & 0.97 \\
Midway anomaly & \num{50000} & 0.001 & 0.99 & 0.99 \\
Late anomaly & \num{99000} & 0.05 & 1.0 & 1.0 \\
\bottomrule
\end{tabular}
\end{table}

\subsubsection{The very early anomaly: $\tau = 1$}

We next consider the case in which the anomaly is the initial vertex of PA network. This setting is similar to the well-known superstar model~\cite{bhamidi2015superstar}, in which the newly added vertex is attached to the superstar with fixed and positive probability. We keep the parameter settings the same as in the experiments where $\delta$ is unknown, with $\delta = 0.1$, $m = 3$, $\alpha = 0.05$, and $t = 10^{4}$, and we select the 5 candidate vertices with the highest degree and 5 vertices with highest ratio as  discussed in Section~\ref{subsec: Detection unknown delta}.

\begin{figure}[htbp]
    \centering
    \includegraphics[width=0.5\linewidth]{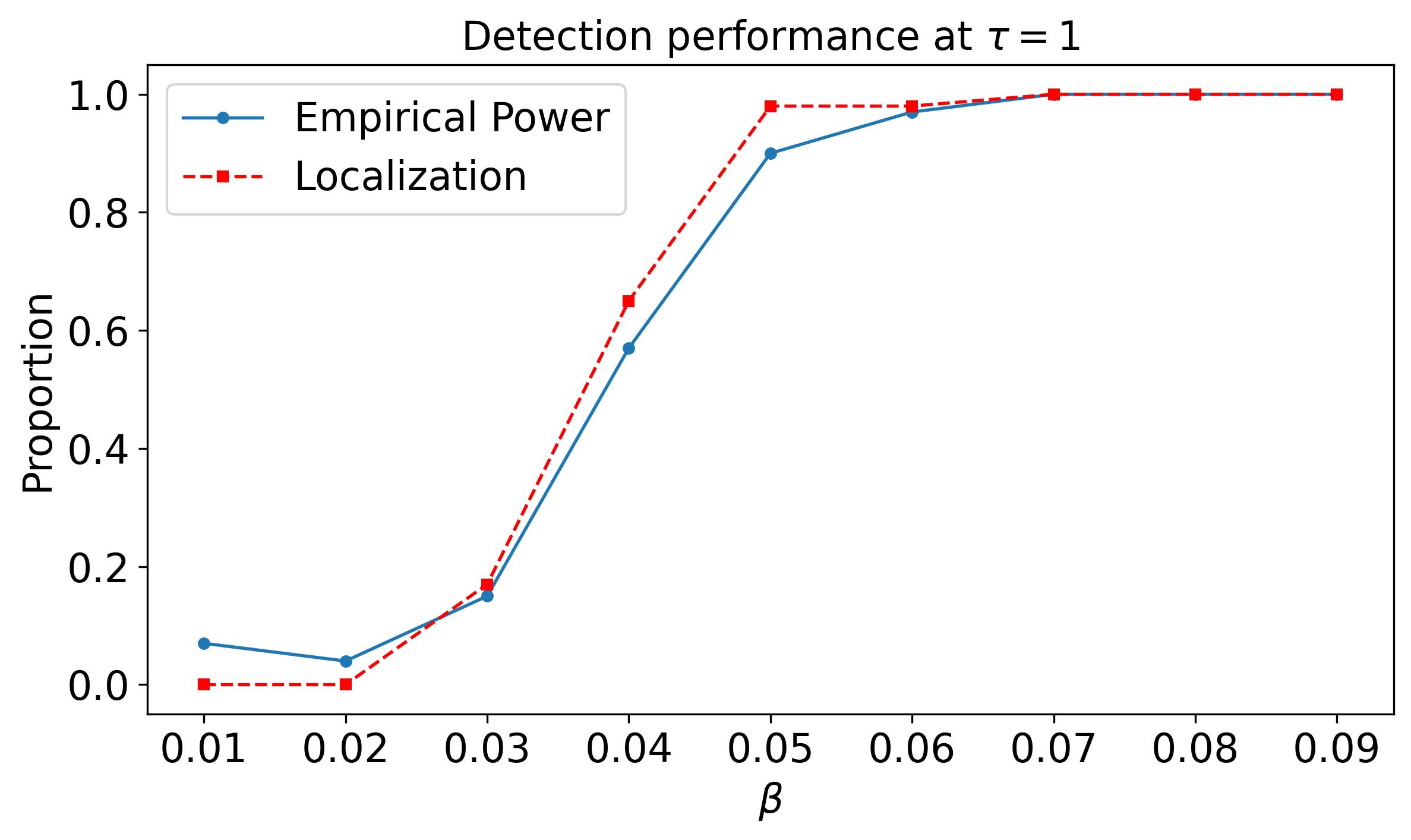}
    \caption{ Detection performance when $\tau = 1$, $t = 10^{4}$, $\delta = 0.1$, and $m = 3$, where $\delta$ is unknown during detection. }
    \label{fig: detection_initial_vertex}
\end{figure}

The results are shown in Figure~\ref{fig: detection_initial_vertex}. As $\beta$ increases, both empirical power and localization rate increases. Nevertheless, when $\beta$ is small, detection becomes less effective. This confirms that the weak anomalous signals are difficult to distinguish from the inherent stochastic variability of early-stage preferential attachment dynamics. 

\subsection{ Interpretation of the results}

In summary, the experiments demonstrate that the estimation and detection procedures perform well across a range of anomaly regimes and parameter settings:
\begin{enumerate}[label=(\arabic*)]
    \item The estimations of $\beta$ and $\delta$ are centered around the true values, and the empirical distributions of $\hat{\beta}$ and $\hat{\delta}$ are similar even when we use different initial values of $\beta$, which means that the joint estimation procedure is insensitive to the initial values.
    \item Both the empirical power and the localization rate increase monotonically with $\beta$ across all anomaly regimes, confirming that stronger anomalies are consistently easier to detect and localize.
    \item Among the three regimes, midway anomalies are the most reliably detected. In contrast, very early and, surprisingly, very late anomalies are more challenging, but for different reasons: early anomalies are affected by high variability in degree growth, while late anomalies have insufficient time to accumulate extra edges.
    \item The choice of candidate vertices plays a crucial role in detection performance. Selecting vertices based on top ratio alone is sufficient for midway and late anomalies, and the combined criterion performs better for more extreme early cases.
\end{enumerate}

\section{Discussion}

In this study, we proposed a comprehensive likelihood-based framework for jointly estimating parameters and detecting anomalies in preferential attachment networks. Initially, the problem may seem easy because the anomaly attracts new edges at linear rate. Our first idea was that the anomaly could be quickly detected based only on its degree growth rate. Surprisingly, due to high variability in the degree evolution in the PA networks, all algorithms based on this simple idea failed even for relatively large $\beta$. On contrary, the proposed approach proved effective across a wide range of settings.

A key strength of the proposed method lies in its temporal adaptability: the detection and localization performance remains stable regardless of whether the anomaly occurs at an early, midway, or late stage of the network evolution. By capturing systematic deviations in the attachment dynamics over time, the framework offers a practical and effective tool for monitoring the structural integrity of evolving networks.

Existing studies on anomaly detection in dynamic networks span a wide range of problem formulations and perspectives. Some likelihood-based studies~\cite{cirkovic2022likelihood, du2025Likelihood, Kaddouri2026Likelihood} focus on change-point detection in PA networks, identifying the structural changes when parameter $\delta$ shifts in PA networks. Our framework operates on one anomaly in the entire PA network which allows us to pinpoint the specific anomalous vertex rather than detecting a temporal deviation in the network evolution. 

Our method has several limitations. First, the theoretical properties of our parameter estimates and test statistics remain unknown. Indeed, the likelihood function becomes analytically intractable once an anomaly occurs, making it challenging to establish the consistency of the MLEs of $\beta$ and $\delta$, and no closed-form estimators are available. As a result, the estimation procedure relies entirely on numerical optimization. Similarly, the statistical significance and the power of the test are assessed through empirical evaluation rather than explicit theoretical guarantees. Second, computing the critical region requires generation and log-likelihood maximization for many synthetic networks. When a network is large, this procedure can be computationally expensive.

\section{Conclusion and future work}

In this research, we proposed an iterative parameter estimation method and a likelihood-based framework to detect anomalies in PA networks. We also explored the strategy to reduce computational costs and evaluated the performance of our method under various anomaly settings and parameter regimes. Building on these findings, we outline several promising directions for future work:
\begin{enumerate}
    \item \textbf{Multiple anomalies.} Our current model assumes the occurrence of a \emph{single} anomaly during network evolution. A natural extension is to generalize the detection framework to handle multiple anomalies simultaneously, potentially with different attachment mechanisms. This would require jointly estimating multiple anomalous vertices and their associated parameters, which could substantially increase the computational complexity of the estimation procedure.
    
    \item \textbf{Extensions of the method.} In this study, the key indicator of an anomaly is the increased number of incoming edges to the anomalous vertex, which reduces the number of edges received by other vertices compared to those in a standard PA network. As a result, minimal degree testing, as proposed in \cite{kay2023detecting}, may serve as an alternative method for anomaly detection in our setting. Another interesting direction is to test whether $\beta$ exceeds a given threshold, which may provide complementary insights into the network growth mechanism. Preliminary results suggest that the performance is comparable to the likelihood ratio procedure presented in this study. A thorough theoretical and empirical investigation is left for future work. Furthermore, other potential approaches worth exploring include using KL divergence to measure distributional differences or leveraging properties of the Laplacian matrix and its eigenvalues~\cite{huang2020Laplacian} for structural anomaly detection.
    
     \item \textbf{Realistic datasets.} All experiments in this study were conducted on synthetically generated PA graphs. An important direction for future work is to evaluate the proposed method on real-world datasets, such as citation networks and social media graphs, to assess its performance in more realistic scenarios.
     
    \item \textbf{Theoretical improvements.} From a theoretical perspective, as previously noted, obtaining a closed-form solution for the MLEs of $\beta$ and $\delta$ is analytically challenging. Consequently, establishing the consistency of these estimators in our model remains an open theoretical problem. Future work should aim to rigorously analyze the asymptotic properties of the proposed estimators.
    \item \textbf{Different anomaly setting.} In our current model, the parameter $\beta$ influences the degree of the anomalous vertex, which in turn affects the probability distribution, and increases the complexity of the likelihood function. Future work could explore alternative anomaly settings that may more significantly alter the overall dynamics of the network evolution, potentially leading to richer models and more robust detection mechanisms, such as considering time-varying anomaly effects, allowing the attachment bias of anomalous vertices to evolve over time.
    \item \textbf{Online and sequential detection.} Our proposed procedure relies on the availability of the entire graph evolution for global analysis. An interesting direction is to develop detection methods that operate in real time as the graph evolves, enabling the identification of anomalies as soon as they occur, as studied in evolving networks~\cite{Peel2015Baysian} and in graph anomaly detection more broadly~\cite{Akoglu2015detection}. Potential approaches include developing incremental versions of the proposed likelihood-based framework that can be updated as new vertices and edges arrive, as well as designing time-sensitive detection statistics to capture early-stage abnormal behaviors.
    \item \textbf{Scalability and computational complexity.} Although we adopt degree-based ranking to narrow the search range, scaling the method to massive networks remains a challenge. This will require approximate inference techniques and more efficient search strategies over the anomaly location. Developing such scalable solutions would broaden the applicability of the proposed method to real-world networks, which are often large and complex.
\end{enumerate}

\paragraph{Acknowledgments.}
The work of QL was supported by the China Scholarship Council. RvdH and NL are partially supported by the Netherlands Organisation for Scientific Research (NWO) through the Gravitation NETWORKS grant 024.002.003.

\bibliographystyle{plain}
\bibliography{references} 

\end{document}